\documentclass[english,svgnames,pre]{revtex4-2}
\usepackage[T1]{fontenc}
\usepackage[latin1]{inputenc}
\usepackage{geometry}
\usepackage{color}
\usepackage{float}
\usepackage{bm}
\usepackage{amsmath}
\usepackage{amssymb}
\usepackage{graphicx}
\usepackage{wasysym}

\makeatletter
\usepackage{lipsum}
\usepackage{color}
\definecolor{yellow}{RGB}{255,255,0}
\makeatother

\usepackage{babel}
\begin{document}
\title{Response of a dusty plasma system to external charge perturbations}
\author{Hitendra Sarkar and Madhurjya P.\ Bora}
\affiliation{Physics Department, Gauhati University, Guwahati 781014, India}
\begin{abstract}
The excitation of nonlinear wave structures in a dusty plasma caused
by a moving external charge perturbation is examined in this work,
which uses a 1-D flux corrected transport simulation. The plasma responds
uniquely to different nature of the moving charge, depending on which,
for small amplitude perturbations, pinned envelope solitons are generated
and electrostatic dispersive ion-acoustic shock waves are formed for
a large amplitude perturbation. The presence of dust particles is
found to suppress the formation of dispersive shocks  at low
velocity of the external charge debris. The results are also investigated
theoretically as a solution to the generalized Gross-Piteavskii equation,
which broadly supports the simulation results. 
\end{abstract}
\maketitle

\section{Introduction}

The formation of nonlinear structures in plasmas is a subject of considerable
interest for researchers due to their complexities and intricate nature.
Apart from being a subject of applied plasma physics, they also generate
a lot of academic interests as we continue to unravel new physical
processes which may generate these structures in plasmas. Numerous
theoretical and experimental studies have been conducted to investigate
the origin, dynamics and properties of such structures forming in
a nonlinear plasma. One such process which excites various nonlinear
structures in plasma is external charge perturbations. These perturbations
can arise due to the presence of external material in a plasma such
as debris in a flowing plasma. A perfect example of this particular
case would be the perturbation due to moving space junks or debris
through the ionosphere. However, various space and astrophysical plasmas,
and even routinely created laboratory plasmas can be relevant in this
context. In this work, we carry out a detailed analysis of different
kinds of such external charge perturbations in a flowing complex plasma
consisting of electrons, ions, and dust particles.

Among various kinds of nonlinear structures in plasmas, solitons or
solitary waves are the most commonly studied structures. A solitary
wave is characterized by its sustained shape over the spatial and
temporal domains without any energy dissipation, which originates
from a balance between the nonlinearity and dispersion in a system.
First studied by Russell in water waves \citep{Russel1844}, solitons
are routinely observed in ion-acoustic (IA) plasma waves. Theoretically,
solitons are found in the form of nonlinear superposition of the waves
as a solution to the Korteweg--De Vries (KdV) equation \citep{Lax1968}.
Another class of nonlinear structures are the envelope solitons characterized
by their slowly varying envelope function which modulates the phase
and amplitude of the carrier wave which moves relatively fast, due
to which envelope solitons can propagate maintaining its constant
shape. These also arise from the balance between dispersion and nonlinearity,
which can be well explained as one of the solutions to the nonlinear
Schr\"odinger equation (NLSE) \citep{Ichikawa1978}. In addition to
solitons, the NLSE and KdV-Burgers equation (KdV-B) can also give
rise to shock waves \citep{Kamchatnov20042,El2009,Kamchatnov2002,Nakamura1999}.
Shock waves are a class of nonlinear structures, which are characterized
by rapid changes in the wave amplitudes and phases and they arise
due to the balance between dispersion and nonlinearity, but unlike
solitons, shocks are formed in the regions where the nonlinearity
is particularly strong. As the nonlinearity dominates over dispersion,
nonlinear effects become more significant and the wave becomes steep,
leading to the formation of a shock \citep{doi:10.1142/4513}. Shocks
can be either monotonic or oscillatory depending upon the dispersion
and unlike the solitons they evolve with time and might breakdown
to form multiple solitons \citep{Nakamura2002,El2016}. We further
note that, these kinds of structures and their studies are not limited
to the domain of plasma physics but also exist in the cases of ocean
waves, Bose-Einstein condensates and nonlinear optics etc., to name
a few \citep{grimshaw1997,Kamchatnov20042,Gedalin1997}. The properties
and the flow dynamics of the nonlinear wave are mostly the same for
all those cases and can be studied using hydrodynamic or fluid equations. 

In recent years, study of plasma flow past an obstacle has been one
of intriguing issues in plasma research, mainly due to its applications
in the space plasma environments. Predictions by theoretical models
on excitation of precursor solitons, pinned solitons and dispersive
shocks using both analytical and forced KdV (f-KdV) equation solution
as well as molecular dynamic simulation \citep{Sen2015,tiwari20162,Tiwari2016},
have been experimentally validated \citep{Jaiswal2016,Arora2019,Arora2021}
for the case of dust-acoustic (DA) waves in a complex plasma. Studies
have observed acceleration or bending of pinned dust-ion-acoustic
solitary waves in both spatially and temporally in
 presence of charged space debris, depending on source debris function
\citep{Acharya2021}. Studies are also done on how the size, shape,
velocity and even the location of existence of the orbital debris
can affect the amplitude, width and frequency of the solitons using
a f-KdV equation and dusty plasma experiments as a detection tool
for collision-free mapping of debris population \citep{Truitt2020,Arora2019}.
Periodic, quasi-periodic and chaotic structures are also studied using
f-KdV and modified KdV model for a Thomas-Fermi plasma, where the
dynamics of the ion-acoustic waves are dependent on the concentration
of positrons, source debris velocity and its strength \citep{Mandi2019}.
Solitary wave solution in $q$-nonextensive plasmas for dust-ion-acoustic
(DIA) waves are also studied using damped forces modified KdV model
\citep{Mandi20192}. Some special exact accelerated soliton solutions
are derived which retain constant amplitudes with varying velocities
over time, whereas the solitonic debris functions can change their
shapes with varying velocities with time in lower-earth orbital (LEO)
region plasma in the presence of charged space debris \citep{Mukherjee2021}.
A realistic case of ion-acoustic solitary waves with damping originated
from orbital debris source has also been explored \citep{Truitt20202}.
Recently a study on moving charge obstacle in a weakly nonlinear and
highly dispersive limit using a driven forced NLSE model, has shown
the formation of collision-less shocks \citep{Chakraborty2022}. 

In this work, we carry out a detailed numerical and theoretical analyses
to investigate how the dynamics of a dusty plasma system is affected
by an external charge perturbation in the ion-acoustic regime, which
has not been explored completely. As dusts are almost native to the
space plasma environments, it is quite important that we know exactly
how their presence effects the plasma response to these perturbations.
We study the response of the plasma to a moving charged perturbation,
in weakly and highly nonlinear regimes by using a 1-D flux-corrected
transport (FCT) simulation, mainly focusing on the
 excitation of envelope solitons and dissipation-less shock waves
(DSWs). Additionally, we have also investigated whether the dust-charge
fluctuation can have any significant impact on the nonlinear IA waves
as it has been shown that dust-charge fluctuation, as a natural phenomenon
associated with most dusty plasma systems, can induce damping to the
IA oscillation, which is also verified theoretically as well as through
simulations \citep{Changmai2020}. We also check if the charging nature
of the external debris (positively charged and negatively charged)
has different influence on the plasma wave and its nonlinear structures,
since a debris can be charged in either way in a space plasma environment,
depending on the material and the environment where they are embedded
\citep{Cui1994}.

\subsection{Dispersive (dissipation-less) shock}

In general, dissipative shocks in non-magnetized plasma are seen in
the weakly nonlinear regime for both ion acoustic and dust-ion acoustic
plasmas and are confirmed by numerical simulation as well as laboratory
experiments \citep{Saitou2003,Bailung2008,Nakamura1999,Nakamura2002}.
This kind of shock structure can originate in collision-less plasmas
due to enhanced Landau damping, theoretically which is solved by modifying
the KdV equation with an extra integral term. It can also arise due
to viscous dissipation or the dissipation of the potential energy
which is generally described by KdV-Burger\textquoteright s equation.
In above cases both oscillatory as well as monotonic shocks are found
and the transition from oscillatory to monotonic shocks are also well
studied \citep{Bailung2008,Nakamura1999,Nakamura2002}. In contrast
to the dissipative shocks, dispersive shock waves (DSW) also known
as dissipation-less shock waves can be viewed as the counterpart of
the well-known viscous shock waves (VSW) or dissipation-led shock
waves. Unlike VSWs, DSWs do not dissipate any energy and the potential
energy discontinuity across a shock front is related to the kinetic
energy associated with the wave modulation while in case of VSWs,
the potential-energy jump is directly due to the dissipation. So,
while the former does not cause an increase in entropy of the system,
the latter does. Structurally, DSWs have oscillatory microstructures
while VSWs are monotonic in nature. In our numerical simulation, the
formation of DSWs is very evident, which we present in this section.
The full model to be solved through the FCT simulation are given by
the Eqs.(\ref{eq:gencont}-\ref{eq:s},\ref{eq:pois-3}) which contain
full dust-charge fluctuation dynamics. Recently a theoretical study
has predicted the excitation of collision-less (dissipation-less)
shock in weakly nonlinear and highly dispersive plasma, arising due
to a highly charged obstacle in a non-viscous, plasma medium moving
with a supersonic velocity using a forced NLSE model (f-NLSE) \citep{Chakraborty2022}.
Another study has also predicted the formation of soliton-like pulses
and dispersive shocks in strongly coupled complex plasmas using molecular
dynamic simulation \citep{Tiwari2016}. However, excitation of dispersive
shocks in an unmagnetized, non-viscous plasma in presence of dust
and the transition from dispersive shocks to envelope solitons are
not systematically studied to the best of our knowledge. Here, in
this paper, we examine how envelope solitons are formed in the near-sonic
region of a moving charge perturbation in a weakly nonlinear regime
and transition of precursor oscillations to a dispersive oscillatory
shock as we move to a highly nonlinear regime. We also study the formation
of pinned envelope solitons and dispersive shocks in ion-acoustic
regime with an admixture of dust as well as in the presence of dust
charge fluctuation. We also comment on how the sign of the charge
of the external perturbation effects the nonlinear waves and formations
of nonlinear structures. We support our simulation findings through
a generalized Gross-Piteavskii equation.

This paper is organized as follows. In section II basic fluid equations
for the plasma model in equilibrium are derived with relevant parameters.
Section III contains the numerical scheme for obtaining the solutions
of the model equations in the form of generalized continuity equations
along with energy conservation for the numerical model in section
III-A and a nonlinear finite difference solver to solve the Poisson
equations explicitly in section III-B. The benchmarking of our numerical
scheme is also provided in section III-C using linear perturbation
method on a dusty plasma with dust charge fluctuation. In section
IV numerical results obtained from external charge perturbation in
both weakly nonlinear and highly nonlinear regimes are presented as
well as their physical and dynamical characteristics are discussed.
In section IV-B we also include a theoretical approach to solve our
problem using f-NLSE equation and draw a parallel between f-NLSE solutions
and our results. Section V shows the comparison of numerical results
for positive and negative charge perturbation. Finally, a brief summary
of the results and concluding remarks are given in Section VI. 

\section{Theoretical formulation}

We consider a one-dimensional three-fluid model of the plasma --
electrons, ions, and dust particles in the ion-acoustic regime. The
effect of dust particles is realized through the Poisson equation
and dust-charge fluctuation equation. The dust density, however, remains
constant which is a reasonable approximation in the ion-acoustic time
scale. The basic equations are
\begin{eqnarray}
\frac{\partial n_{i}}{\partial t}+\frac{\partial}{\partial x}(n_{i}v_{i}) & = & 0,\label{eq:cont-1}\\
m_{i}n_{i}\frac{dv_{i}}{dt} & = & -\frac{\partial p_{i}}{\partial x}-en_{i}\frac{\partial\phi}{\partial x},\\
\epsilon_{0}\frac{\partial^{2}\phi}{\partial x^{2}} & = & e(n_{e}-n_{i})-q_{d}n_{d},
\end{eqnarray}
where the symbols have their usual meanings and $q_{d}$ is the dust
charge. The electrons are assumed to be Boltzmannian. We now normalize
the densities by their respective equilibrium values i.e.\ $n_{j}\to n_{j}/n_{j0}$,
where the subscript `0' refers to the equilibrium values and $j=e,i,d$
respectively for electrons, ions, and dust particles. The ion velocity
$v_{i}$ is normalised with the ion-sound velocity $c_{s}=\sqrt{T_{e}/m_{i}}$,
where $T_{i,e}$ are the ion and electron temperatures, measured in
the units of energy and are held constant. The length is normalised
with the electron Debye length $\lambda_{D}$ and time is normalised
with the \emph{zero-dust} ion-plasma frequency 
\begin{equation}
\omega_{pi}=\left(\frac{n_{0}e^{2}}{m_{i}\epsilon_{0}}\right)^{1/2},
\end{equation}
where $n_{0}=n_{e0}=n_{i0}$. The potential $\phi$ is normalised
with $(T_{e}/e)$. The ion thermal pressure is given by $p_{i}=n_{i}T_{i}$,.

We shall assume that the dust particles are negatively charged and
acquire negative potential with respect to the bulk plasma
\begin{equation}
q_{d}=-ez_{d},\label{eq:q}
\end{equation}
where $z_{d}$ is the dust-charge number, so that the above equations
in a dimensionless form can be written as
\begin{eqnarray}
\frac{\partial n_{i}}{\partial t}+\frac{\partial}{\partial x}(n_{i}v_{i}) & = & 0,\label{eq:cont}\\
n_{i}\frac{dv_{i}}{dt} & = & -\sigma\frac{\partial n_{i}}{\partial x}-n_{i}\frac{\partial\phi}{\partial x},\label{eq:mom}\\
\frac{\partial^{2}\phi}{\partial x^{2}} & = & n_{e}-\delta_{i}n_{i}+\delta_{d}z_{d},\label{eq:pois}
\end{eqnarray}
where $\sigma=T_{i}/T_{e}$, $\delta_{i}=n_{i0}/n_{e0}$, and $\delta_{d}=n_{d}z_{d0}/n_{e0}$.
The dynamic dust-charge number $z_{d}$ is normalised with its equilibrium
value $z_{d0}=z_{d}|_{\phi=0}$. We note that the quasi-neutrality
condition can now be expressed as $\delta_{i}=1+\delta_{d}$. The
dust-charge $q_{d}$ for spherical dust particles can also be expressed
in terms of dust potential $\varphi_{d}$
\begin{equation}
q_{d}=C\,\Delta V=4\pi\epsilon_{0}r_{d}\varphi_{d},
\end{equation}
where $C$ is the grain capacitance and $\varphi_{d}=\phi_{g}-\phi$,
$\phi_{g}$ being the grain potential. We define the equilibrium dust-charge
number $z_{d0}$ in terms of the magnitude of the equilibrium dust
potential $\varphi_{d0}=\left|\varphi_{d}|_{\phi=0}\right|$,
\begin{equation}
z_{d0}=4\pi\epsilon_{0}r_{d}e^{-1}\varphi_{d0},\label{eq:fid0}
\end{equation}
where $e$ is the magnitude of electronic charge and $r_{d}$ is the
radius of a dust-particle. By using the relation (\ref{eq:q}), we
have the normalised expression for dust potential
\begin{equation}
\varphi_{d}=-\alpha^{-1}z_{d},
\end{equation}
where the dust potential is normalised by the magnitude of its equilibrium
value $\varphi_{d0}$, defined above and $\alpha=(e\varphi_{d0}/T_{e})^{-1}$
is the ratio (magnitude) of the electron thermal energy to the equilibrium
dust potential energy. 

Let us now consider the dust-charging equation
\begin{equation}
\frac{dq_{d}}{dt}=I_{e}+I_{i},\label{eq:dust-1}
\end{equation}
where $I_{e,i}$ are the electron and ion currents to the dust particles
which can be written as (dimensional) 
\begin{eqnarray}
I_{i} & = & 4\pi r_{d}^{2}en_{i}\left(\frac{T_{i}}{2\pi m_{i}}\right)^{1/2}\left(1-\frac{e\varphi_{d}}{T_{i}}\right),\label{eq:ii}\\
I_{e} & = & -4\pi r_{d}^{2}en_{e}\left(\frac{T_{e}}{2\pi m_{e}}\right)^{1/2}\exp\left(\frac{e\varphi_{d}}{T_{e}}\right).\label{eq:ie}
\end{eqnarray}
Assuming Boltzmannian electron density
\begin{equation}
n_{e}=e^{\phi},
\end{equation}
the normalised dust-charging equation Eq.(\ref{eq:dust-1}) can be
written as
\begin{equation}
\frac{d\varphi_{d}}{dt}=\hat{I}_{e0}\left[\delta_{i}\delta_{m}\sigma^{1/2}n_{i}\left(1-\frac{\varphi_{d}}{\sigma}\right)-\exp(\phi+\varphi_{d})\right],\label{eq:charging}
\end{equation}
where $\hat{I}_{e0}$ is the normalised equilibrium electron current
to the dust particles 
\begin{equation}
\hat{I}_{e0}=\frac{r_{d}e^{2}n_{e0}}{\epsilon_{o}T_{e}\omega_{pi}}\left(\frac{T_{e}}{2\pi m_{e}}\right)^{1/2}
\end{equation}
and $\delta_{m}=\sqrt{m_{e}/m_{i}}\approx0.023$. 

\section{The numerical scheme}

We employ here 1-D flux-corrected transport (FCT) method for numerical
solution of our model equations. The principle of FCT method is to
contain the numerical diffusion by enforcing flux conservation across
the numerical cell interface at every time step. The numerical diffusion
is inherent to finite difference method which arises due to discretization
of the differential operators on a grid. In this work, we specifically
use Zalesak's FCT scheme \citep{Zalesak1979} which
uses a different limiter from Boris's original scheme \citep{Boris1973}
to limit the fluxes.

In order to facilitate the numerical simulation through the FCT formalism,
we should put all the equations, except the Poisson equation, in the
form a generalized continuity equation,
\begin{equation}
\frac{\partial f}{\partial t}=-\frac{\partial}{\partial x}(fv)+c\frac{\partial s}{\partial x},
\end{equation}
where $f$ is any physical quantity, which we want to solve for, $(fv)$
is its flux and the last term of the above equation is the source
term with $c$ as the coefficient of the source term. Using Eq.(\ref{eq:cont}),
Eq.(\ref{eq:mom}) now can be written as
\begin{equation}
\frac{\partial}{\partial t}(n_{i}v_{i})=-\frac{\partial}{\partial x}(n_{i}v_{i}v_{i})+S_{i}.
\end{equation}
The continuity equation Eq.(\ref{eq:cont}) is already in the desired
form
\begin{equation}
\frac{\partial n_{i}}{\partial t}=-\frac{\partial}{\partial x}(n_{i}v_{i}).\label{eq:cont2}
\end{equation}
The dust-charging equation can be multiplied by $n_{i}$ and can be
recast after applying the continuity equation Eq.(\ref{eq:cont2})
\[
\frac{\partial}{\partial t}(n_{i}\varphi_{d})=-\frac{\partial}{\partial x}(n_{i}\varphi_{d}v_{i})+S_{d}.
\]
In the above equations $S_{i,d}$ are respective source terms
\begin{eqnarray}
{\normalcolor S_{i}} & {\normalcolor =} & {\normalcolor -\sigma\frac{\partial n_{i}}{\partial x}-n_{i}\frac{\partial\phi}{\partial x},}\\
{\normalcolor S_{d}} & {\normalcolor =} & {\normalcolor n_{i}\hat{I}_{e0}\left[\delta_{i}\delta_{m}\sigma^{1/2}n_{i}\left(1-\frac{\varphi_{d}}{\sigma}\right)-\exp(\phi+\varphi_{d})\right].}
\end{eqnarray}
So, the whole model now can be framed as a set of 1-D generalized
continuity equations
\begin{equation}
\frac{\partial\bm{f}}{\partial t}=-\frac{\partial}{\partial x}(\bm{f}v)+\bm{S},\label{eq:gencont}
\end{equation}
where
\begin{eqnarray}
\bm{f} & = & (n_{i},n_{i}v_{i},n_{i}\varphi_{d})',\label{eq:f}\\
\bm{S} & = & \left(0,S_{i},S_{d}\right)'.\label{eq:s}
\end{eqnarray}
are 2-D column vectors with $\bm{S}$ representing the source term.
Along with the above equations, the Poisson's equation (\ref{eq:pois}),
which needs to be solved is
\begin{equation}
\frac{\partial^{2}\phi}{\partial x^{2}}=e^{\phi}-\delta_{i}n_{i}-\delta_{d}\varphi_{d}\equiv f(n_{i},\phi,\varphi_{d}).\label{eq:pois-1}
\end{equation}

\subsection{Energy conservation}

While our three-fluid plasma model is quite complete so long as we
do not allow the temperature to change, in order to ensure total energy
conservation of the numerical model, we determine the pressure through
the two-fluid energy conservation equation
\begin{equation}
\frac{3}{2}\frac{\partial p}{\partial t}=-\frac{\partial}{\partial x}\left(\frac{5}{2}pv\right)+v\frac{\partial p}{\partial x},
\end{equation}
which can be written in a generalized continuity equation form as
\begin{equation}
\frac{\partial p}{\partial t}=-\frac{\partial}{\partial x}(pv)+S_{p},
\end{equation}
where
\begin{equation}
S_{p}=-\frac{2}{3}p\frac{\partial v}{\partial x}
\end{equation}
is the equivalent source term.

\subsection{Poisson solver}

We have constructed a nonlinear finite-difference Poisson solver so
as to keep the total plasma potential $\phi$ to zero over the whole
simulation domain $[a,b]$
\begin{equation}
\int_{a}^{b}\phi\,dx=0.
\end{equation}
This helps us maintain the reference potential as zero all the time,
which is quite justified as the charge perturbation (due to debris)
can be considered to be quite small compared to total electron and
ion charge densities. This can be accomplished with the introduction
of a Lagrange multiplier ${\cal L}$ to the Poisson equation Eq.(\ref{eq:pois-1})
\begin{equation}
\frac{\partial^{2}\phi}{\partial x^{2}}+{\cal L}=f(n_{i},\phi,\varphi_{d}),
\end{equation}
with the solvability condition ${\cal L}=0$. The resultant nonlinear
finite-difference problem is solved with the help of Newton's iterations.
We have used two different methods for the linear solver --- LU decomposition
and a linear bi-conjugate method. The simulation code can automatically
switch between the LU decomposition and bi-conjugate method depending
on the simulation grid size as for grid size more than $\apprge150$,
the bi-conjugate method becomes faster than LU decomposition.

\begin{figure}[t]
\begin{centering}
\includegraphics[width=0.5\textwidth]{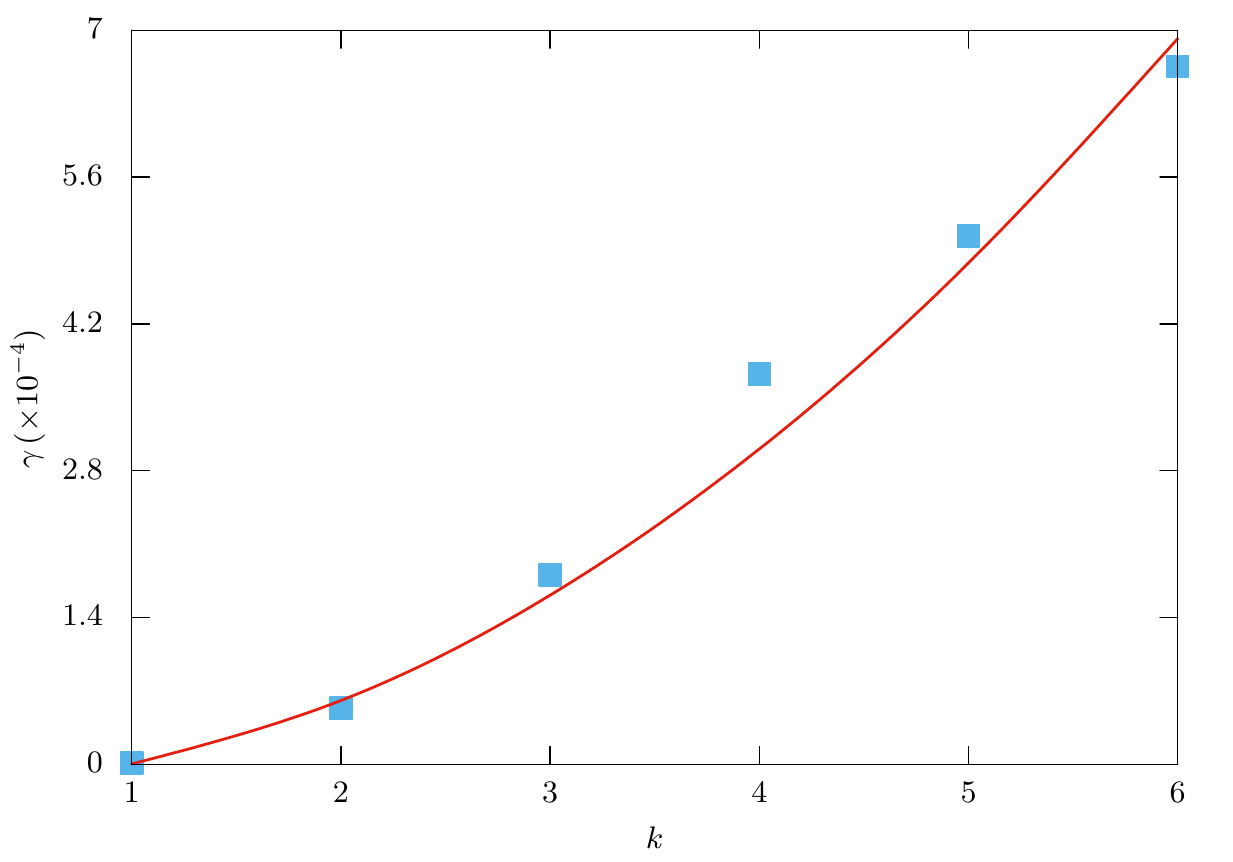}\hfill{}\includegraphics[width=0.5\textwidth]{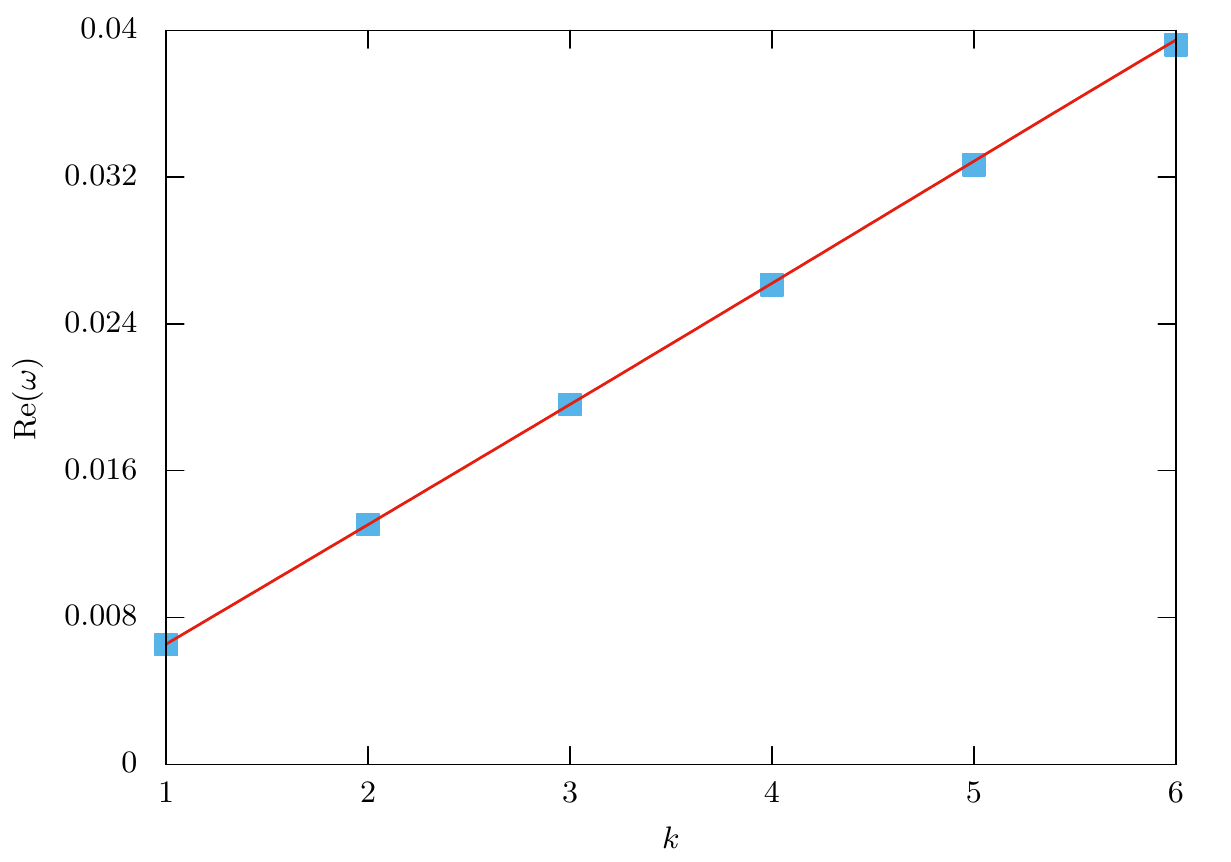}
\par\end{centering}
\caption{\label{fig:The-damping-of}The damping of the ion-acoustic wave due
to dust-charge fluctuation. In the left panel, the normalised damping
rate $\gamma$ is shown against normalised wave number $k$ and in
the right panel, the corresponding frequency of oscillation $\textrm{Re}(\omega)$
is shown. In both the panels the solid line indicates the theoretical
curve as obtained by solving Eq.(\ref{eq:dis}) and the solid square
points indicates the points from the simulation (for relevant plasma
parameters, please see the accompanying text).}
\end{figure}

\subsection{Linear perturbation}

Although the original FCT method does not solve any equations other
than the generalized continuity equations, a two-fluid plasma model
necessarily requires solution of the Poisson equation in space apart
from the other fluid continuity equations. In this context, it is
instructive to examine the response of our numerical model to a linear
density perturbation. The linearised perturbed equations are
\begin{eqnarray}
\frac{\partial n_{i1}}{\partial t}+\frac{\partial v_{i1}}{\partial x} & = & 0,\label{eq:cont1}\\
\frac{\partial v_{i1}}{\partial t} & = & -\sigma\frac{\partial n_{i1}}{\partial x}-\frac{\partial\phi_{1}}{\partial x},\label{eq:mom1}\\
\frac{\partial^{2}\phi_{1}}{\partial x^{2}} & = & \phi_{1}-\delta_{i}n_{i1}-\alpha\delta_{d}\varphi_{d1},\label{eq:pois1}
\end{eqnarray}
where the subscript `$1$' refers to the first order perturbed quantities.
The perturbed dust-charge equation can be written as
\begin{equation}
\frac{\partial\varphi_{d1}}{\partial t}=-\eta\varphi_{d1}-\hat{I}_{e0}\phi_{1}+\hat{I}_{e0}\beta n_{i1},\label{eq:charging-1-1}
\end{equation}
where
\begin{eqnarray}
{\normalcolor \eta} & {\normalcolor =} & {\normalcolor \hat{I}_{e0}\left(1+\delta_{i}\delta_{m}/\sigma^{1/2}\right),}\\
{\normalcolor \beta} & {\normalcolor =} & {\normalcolor \delta_{i}\delta_{m}\sigma^{1/2}\left(1-\frac{\varphi_{d0}}{\sigma}\right).}
\end{eqnarray}
Along with Eq.(\ref{eq:charging-1-1}), the above equations provide
the linear perturbation theory of ion-acoustic wave affected by the
presence of dust-charge fluctuation. Assuming the plasma perturbation
to be of the form $\sim e^{-i\omega t+ikx}$, we obtain the familiar
ion-acoustic wave in absence of any dust effects
\begin{equation}
\omega_{\textrm{IA}}\equiv\omega=k\left(\frac{1}{1+k^{2}}+\sigma\right)^{1/2}.
\end{equation}
If we consider the dust particles without any dust-charge fluctuation,
we get the same dispersion relation as above but with an enhanced
omega for lower $k$
\begin{equation}
\omega_{\textrm{DIA}}\equiv\omega=k\left(\frac{\delta_{i}}{1+k^{2}}+\sigma\right)^{1/2}.\label{eq:dia}
\end{equation}
Note that with dust particles present, $\delta_{i}\geq1$. With dust-charge
fluctuation, the dispersion relation becomes a complex cubic
\begin{equation}
\omega^{3}+i\left(\eta+\xi\right)\omega^{2}-\omega_{\textrm{DIA}}^{2}\omega-i\left(\eta\omega_{\textrm{DIA}}^{2}+\xi\omega_{s}^{2}\right)=0,\label{eq:dis}
\end{equation}
where
\begin{eqnarray}
\omega_{s} & = & k(\sigma+\beta)^{1/2},\\
\xi & = & \frac{\alpha\hat{I}_{e0}\delta_{d}}{1+k^{2}}.
\end{eqnarray}
The above equation admits complex solutions with negative imaginary
part for small $k$, signifying damping of the ion-acoustic wave with
sufficiently long wavelength.

In Fig.\ref{fig:The-damping-of}, we have shown the results of the
simulation in the linear regime. The relevant physical parameters
are $\sigma=0.01,\delta_{i}=1.2,\hat{I}_{e0}\simeq0.17$, which corresponds
to an $e$-$i$ plasma with $n_{i}\sim10^{16}/\textrm{m}^{3},n_{d}\sim4\times10^{12}/\textrm{m}^{3},m_{i}=1.67\times10^{-27}\,\textrm{kg},T_{e}\sim1\,\textrm{eV}$.

\begin{figure}[t]
\begin{centering}
\includegraphics[width=0.5\textwidth]{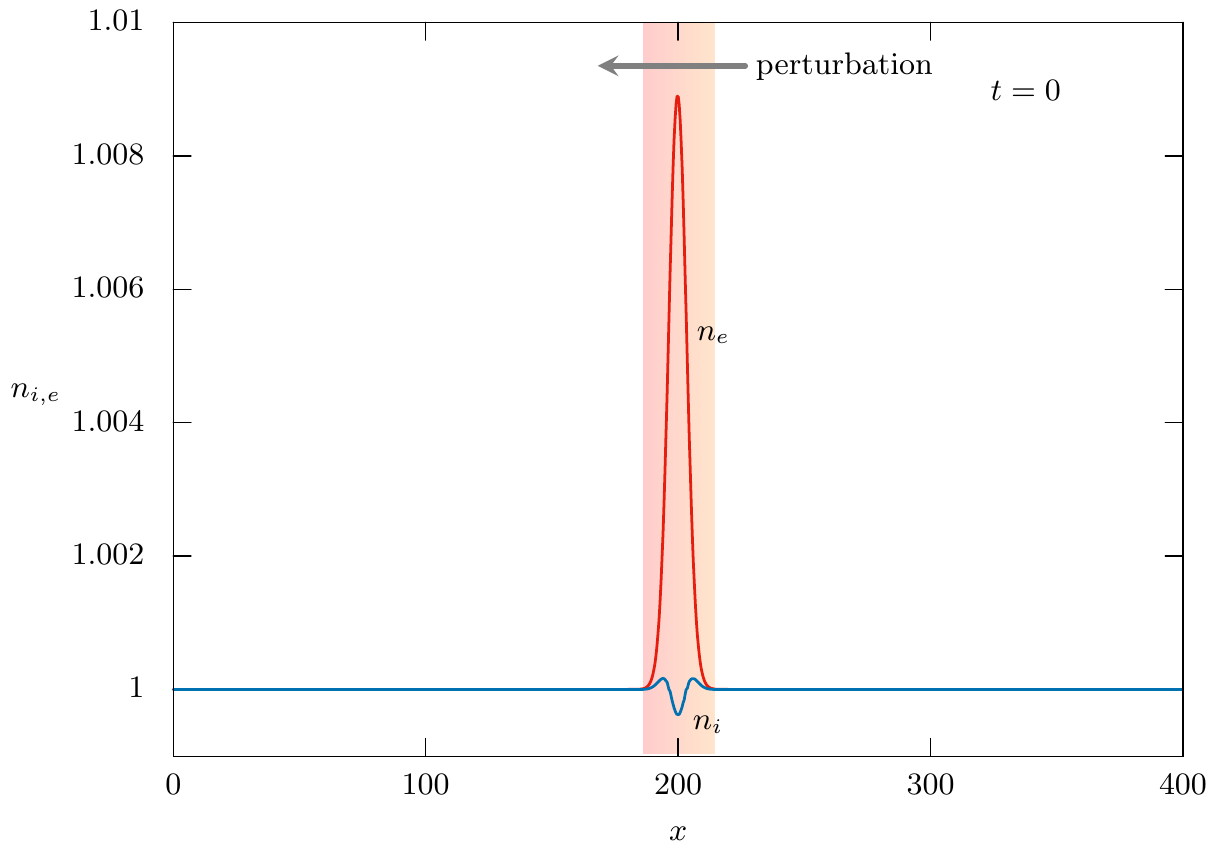}\hfill{}\includegraphics[width=0.5\textwidth]{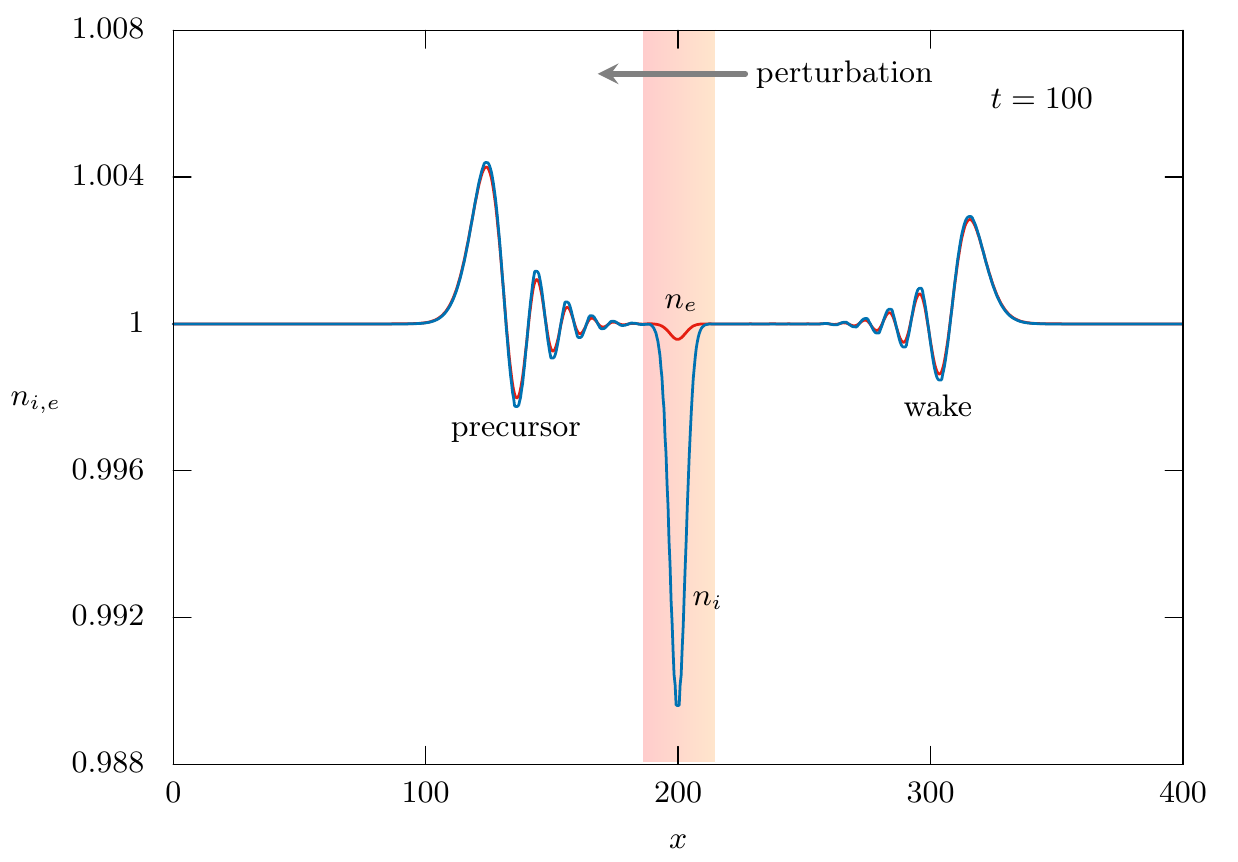}
\par\end{centering}
\caption{\label{fig:Response-of-the}Response of the electron and ion density
to a positive charge perturbation at the initial phase and after about
$100$ plasma periods.}
\end{figure}

\section{External charge perturbation}

We now consider an \emph{unmodulated} external charge
perturbation to the plasma. This can be modeled with an external source
term in the Poisson equation
\begin{equation}
\epsilon_{0}\frac{\partial^{2}\phi}{\partial x^{2}}=e(n_{e}-n_{i})-q_{d}n_{d}+\rho_{{\rm ext}}(x-v_{s}t),\label{eq:pois-3}
\end{equation}
where $\rho_{{\rm ext}}$ is the external charge density moving with
a velocity $v_{s}$. We note that $\rho_{{\rm ext}}$ can either be
positive or negative depending on the charge perturbation, which can
be thought to be a moving charge debris in a plasma. We should note
that as we are considering a complex plasma with dust-charge fluctuation,
the response of the plasma to an external charge perturbation will
be different depending on the nature of the charge. Besides, the morphology
of an external charge perturbation is fundamentally different from
an density perturbation arising out of fluctuation. In the latter,
the response of the plasma may be completely quasi-neutral if the
fluctuations are low frequency perturbations, so that the so-called
\emph{plasma approximation} is automatically satisfied, while ions
will be almost stationary if the fluctuations are high frequency perturbations.

\begin{figure}[t]
\begin{centering}
\includegraphics[width=0.5\textwidth]{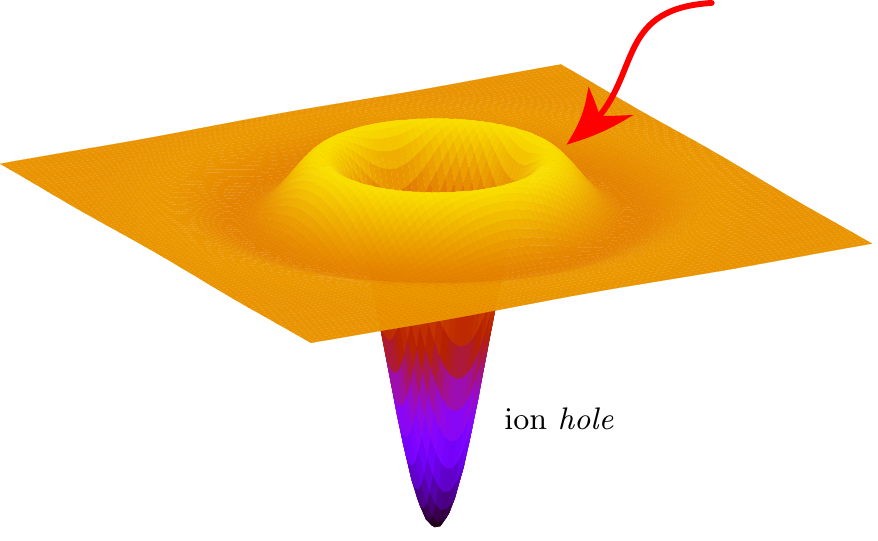}
\par\end{centering}
\caption{\label{fig:A-schematic-representation}A schematic representation
of the response of the plasma to a positive external charge perturbation,
which creates an \emph{ion hole}. The arrow shows the ion density
built-up around the ion hole which disperses as an ion-acoustic wave.}
\end{figure}

In contrast to the above, in case of an \emph{unmodulated} external
charge perturbation, in the vicinity of the perturbation, the response
is always \emph{non-quasi-neutral} as the perturbation never fluctuates,
rather remains embedded in the plasma. However, as the disturbance
due to the perturbation travels with acoustic velocity, away from
the perturbation, the response should be quasi-neutral in nature and
both electron and ion densities should follow each other. This can
be seen form the ion and electron density plots, shown in Fig.\ref{fig:Response-of-the},
arising out of the FCT solver of our model equations with external
charge perturbation, where we have shown the density plots in the
presence of an external positive charge perturbation at the beginning
$(t=0)$ and after $100$ plasma periods when the disturbance has
propagated away from the site of perturbation to the precursor and
wake regions. We have used a Gaussian pulse for the external perturbation
\begin{equation}
\rho_{{\rm ext}}=\hat{\rho}e^{-x^{2}/\Delta},\label{eq:pulse}
\end{equation}
where $\hat{\rho}$ is the perturbation amplitude and $\Delta$ is
a measure of the width of the pulse. The external perturbation The
site of the perturbation is shown as the orange-coloured vertical
strip. Initially the response of the electrons are instantaneous due
their Boltzmannian nature and the electron density peaks up around
the perturbation, similar to a Debye shielding. Due to their inertia,
the ions begin to get depleted slowly and creates a ion-deficient
region (density dip) at the perturbation site. For the same reason,
there is also a built-up of ion density in the vicinity of the perturbation
site as shown in a 3-D schematic representation in Fig.\ref{fig:A-schematic-representation},
Over time, IA oscillations are induced in the plasma as the built-ups
propagates away from the perturbation site as an acoustic wave, which
lead to the creations of the precursors and wakes. As time progresses,
the electron finally responds to the ions and the acoustic disturbance
which is created by the external perturbation and equilibrates with
the ion density everywhere except in the perturbation site, where
the response remains non-quasi-neutral.

In what follows, we first consider the formation of various nonlinear
waves, especially the dispersive shocks and envelope solitons, due
to the external charge perturbation through through the FCT simulation
and then with the well-known reductive perturbative method.

\subsection{Dispersive shock and numerical results}

As we perceive, any low-frequency charge perturbation in a plasma
is supposed to propagate at the ion-acoustic speed throughput the
plasma. The case of an external charge perturbation is no different.
In case of a positive charge perturbation, a localised ion depletion
(or ion hole) occurs with accumulation of ions at the boundary (see
Fig.\ref{fig:A-schematic-representation} for example) due to ion
inertia which subsequently propagates with ion-acoustic speed, which
manifests as precursors and wakes. In the subsequent subsections,
we describe exactly what happens under different circumstances. We
consider the charge perturbation (or the debris) to be moving to the
left (in all figures) at a velocity $v_{s}$, which normalized by
the ion-acoustic speed and can be considered as the Mach number of
the perturbation.

\begin{figure}[t]
\begin{centering}
\includegraphics[width=0.5\textwidth]{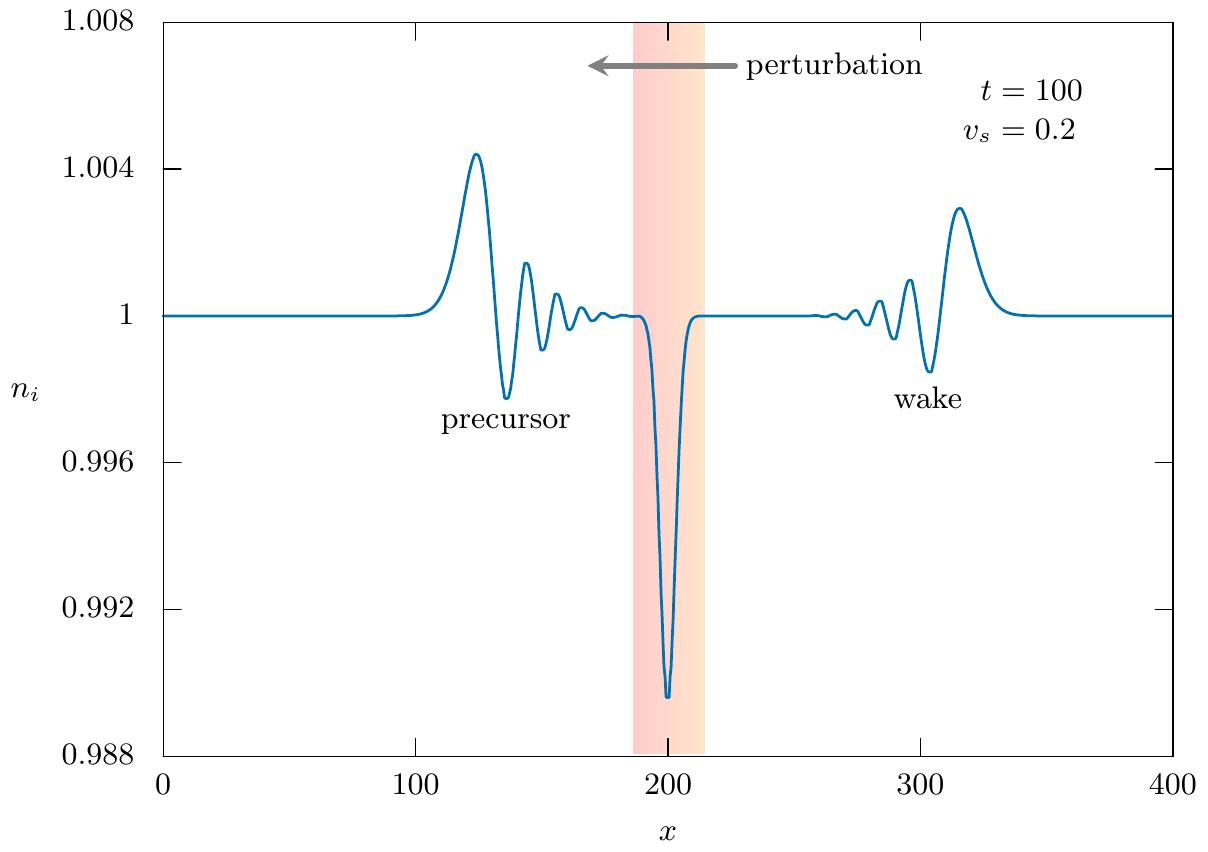}\includegraphics[width=0.5\textwidth]{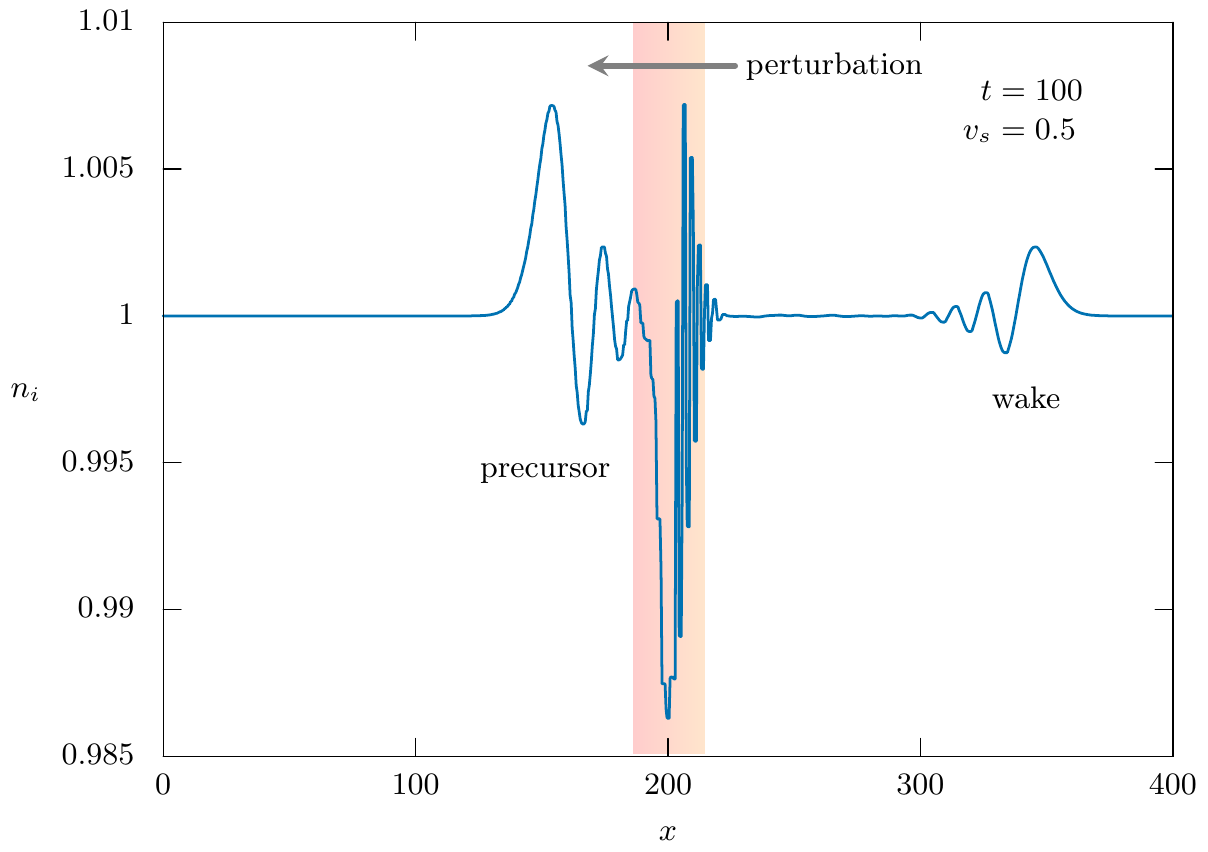}
\par\end{centering}
\begin{centering}
\includegraphics[width=0.5\textwidth]{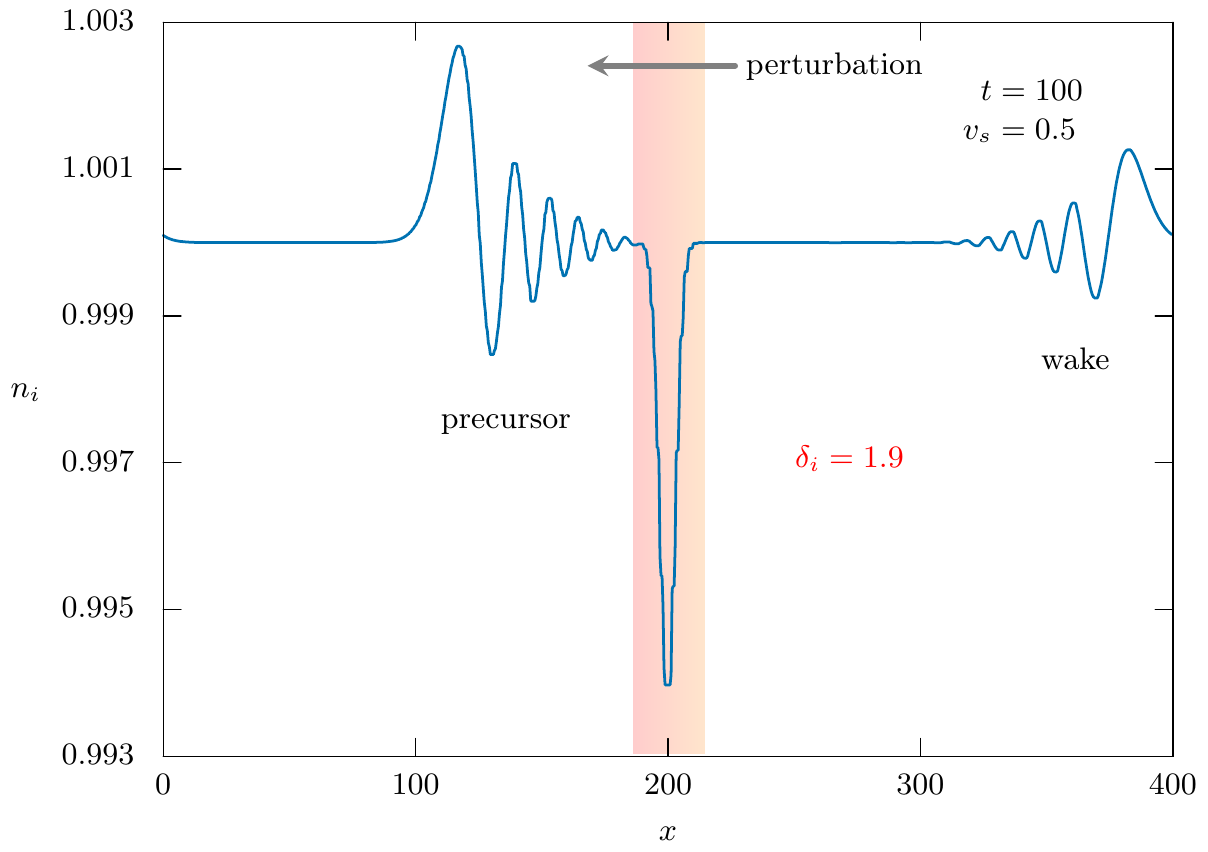}\hfill{}\includegraphics[width=0.48\textwidth,height=6.2cm]{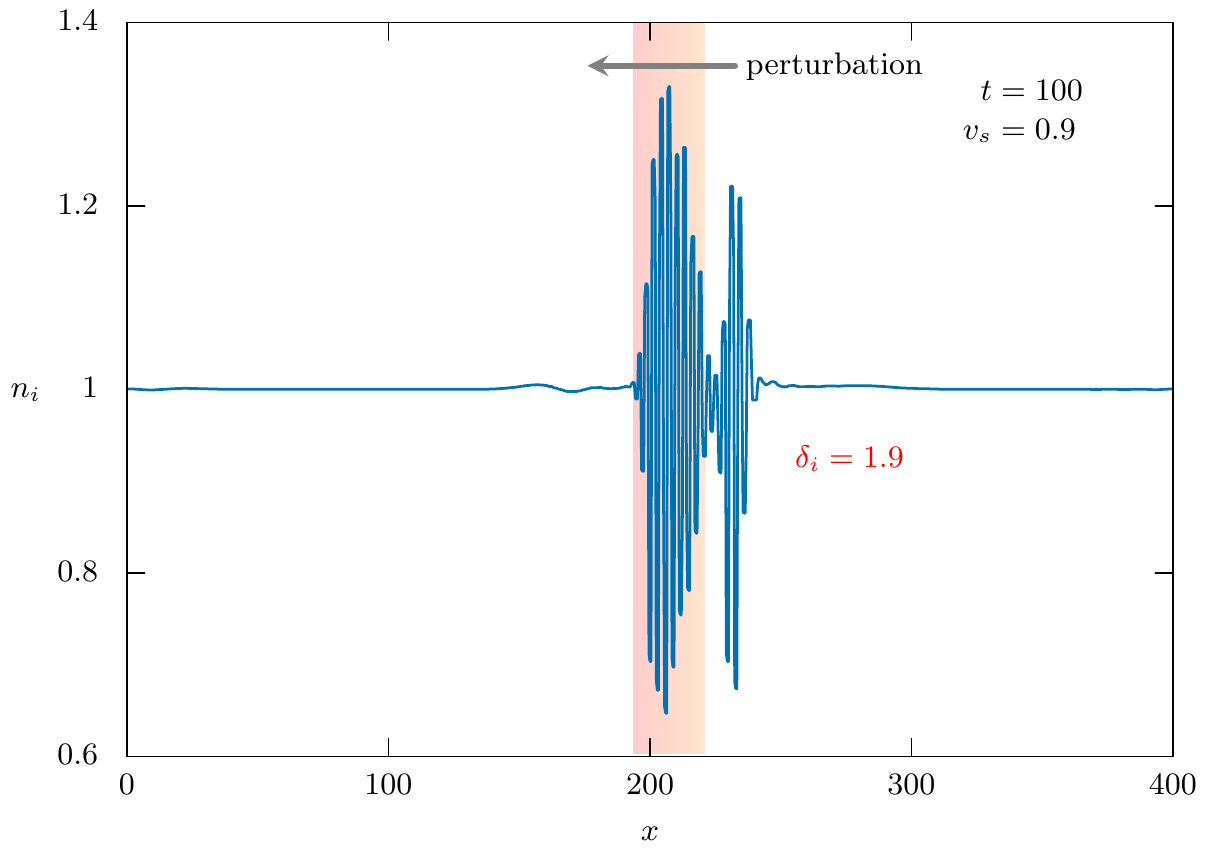}
\par\end{centering}
\caption{\label{fig:The-precursor,-wakes,}The precursor, wakes, and formation
of pinned solitons for low and positive charge perturbation moving
with velocity $v_{s}$ to the left. The bottom panel displays the
effects in the presence of dust. One can compare the 1st figure in
the bottom panel to the 2nd figure of the first panel, both of which
are exactly for same parameters except that the former is with dust
particles.}
\end{figure}

\subsubsection{Effect of dust}

We note that the ion-acoustic wave gets modified to what is known
as dust ion-acoustic (DIA) wave in presence of dust particles 
as shown in Eq.(\ref{eq:dia}) for negatively charged dust particles \citep{Shukla_Mamun_2015}.
In the limit of low frequency perturbation $(k^{2}\lambda_{D}^{2}\ll1)$
and very low ion temperature $(\sigma\ll1)$, the DIA dispersion relation
becomes
\begin{equation}
\omega\simeq k\left(\frac{n_{i0}}{n_{e0}}\right)^{1/2}c_{s}.\label{eq:ia}
\end{equation}
This can be viewed in the present context as an effective increase
in the sound speed with $c_{\textrm{effective}}\equiv c_{s}(n_{i0}/n_{e0})^{1/2}$
as $\delta_{i}=(n_{i0}/n_{e0})>1$. This effectively makes a near-sonic
perturbation subsonic. So, we expect that the effect of velocity of
the perturbation $v_{s}$ should be reduced in presence of dust particles.
However, as $\delta_{i}$ is never very larger than unity, when $v_{s}$
becomes considerably larger than $1$, we \emph{do not} expect the
effect of dust particles to be very pronounced.

\begin{figure}[t]
\begin{centering}
\includegraphics[width=0.5\textwidth]{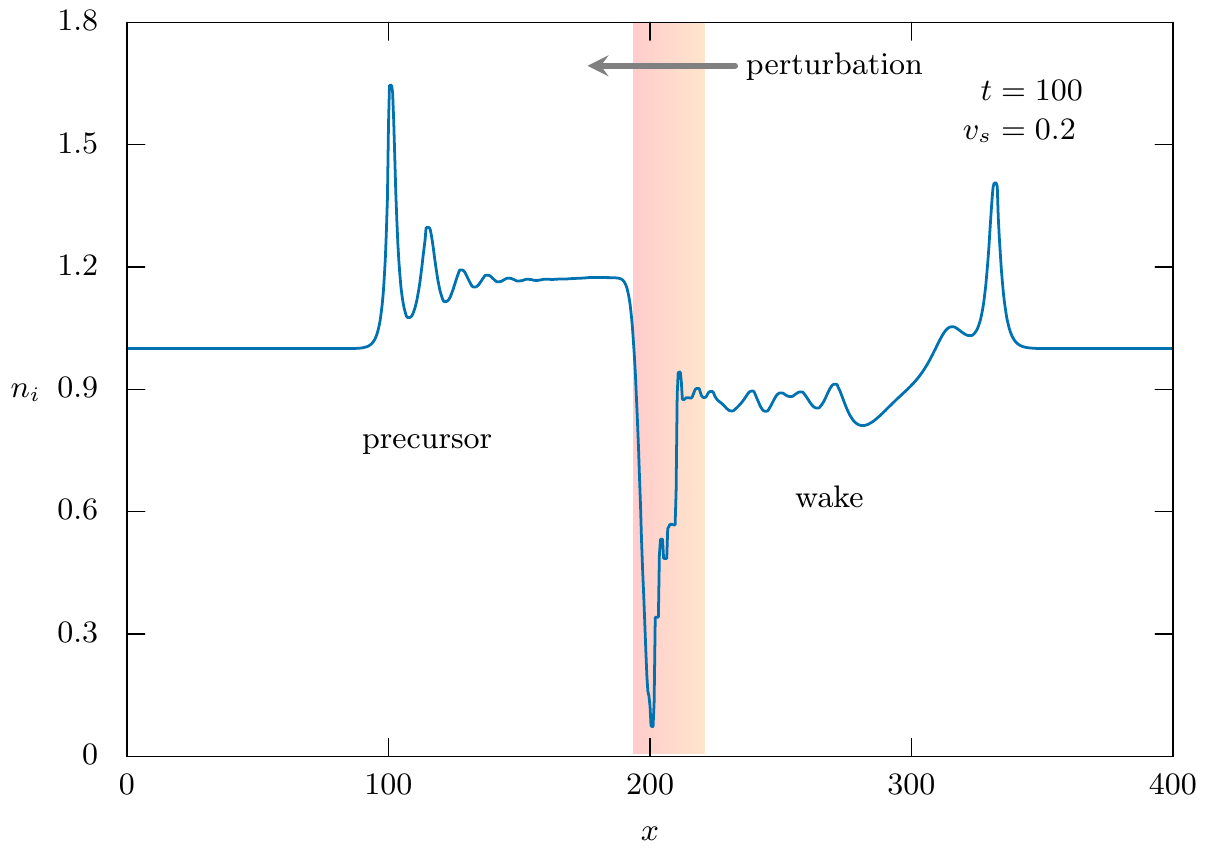}\includegraphics[width=0.5\textwidth]{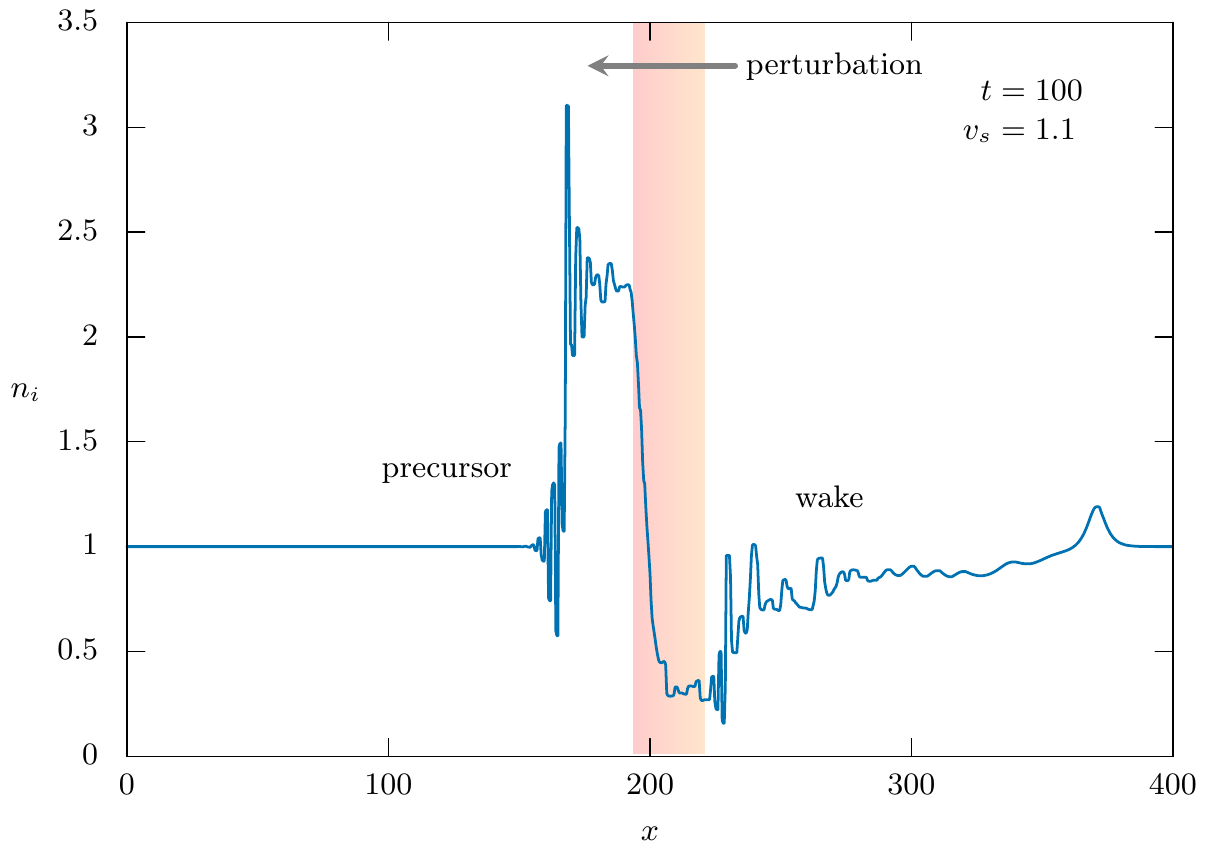}
\par\end{centering}
\begin{centering}
\includegraphics[width=0.5\textwidth]{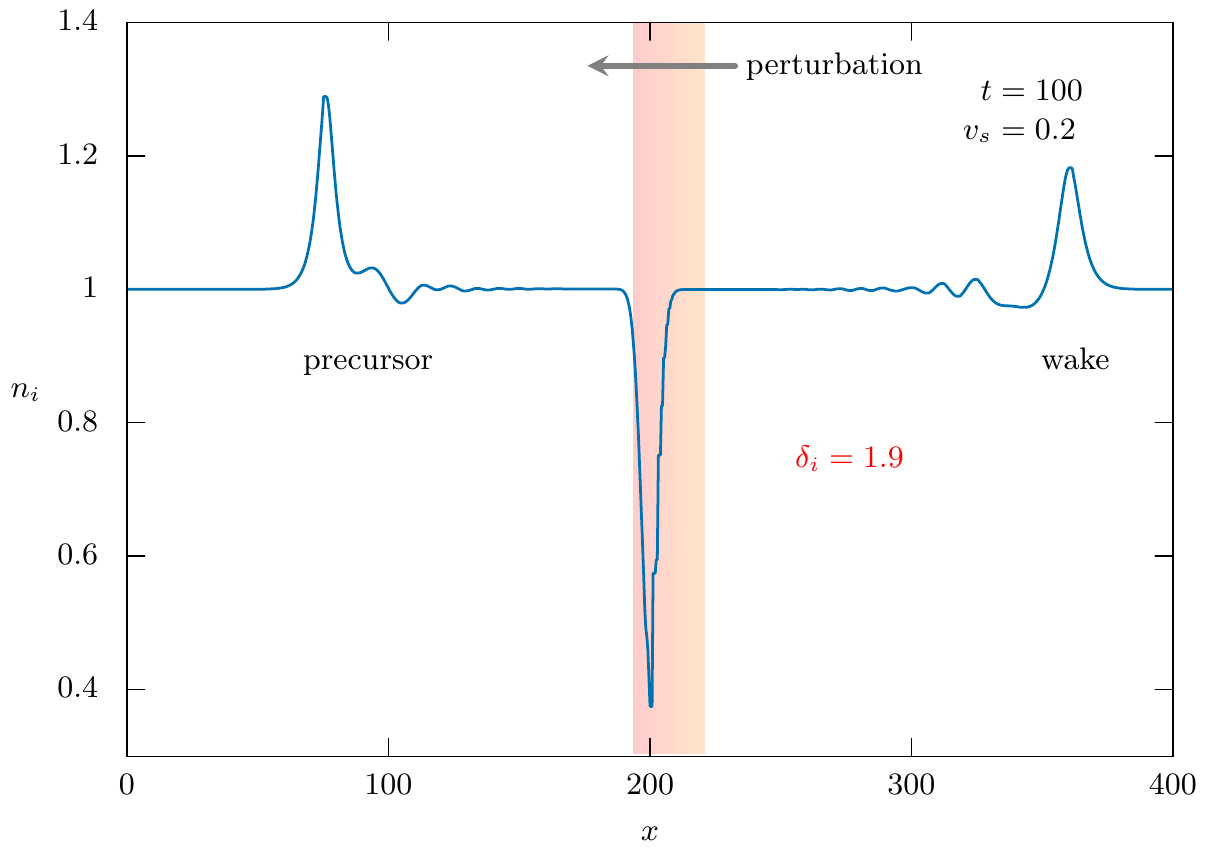}\includegraphics[width=0.5\textwidth,height=6.2cm]{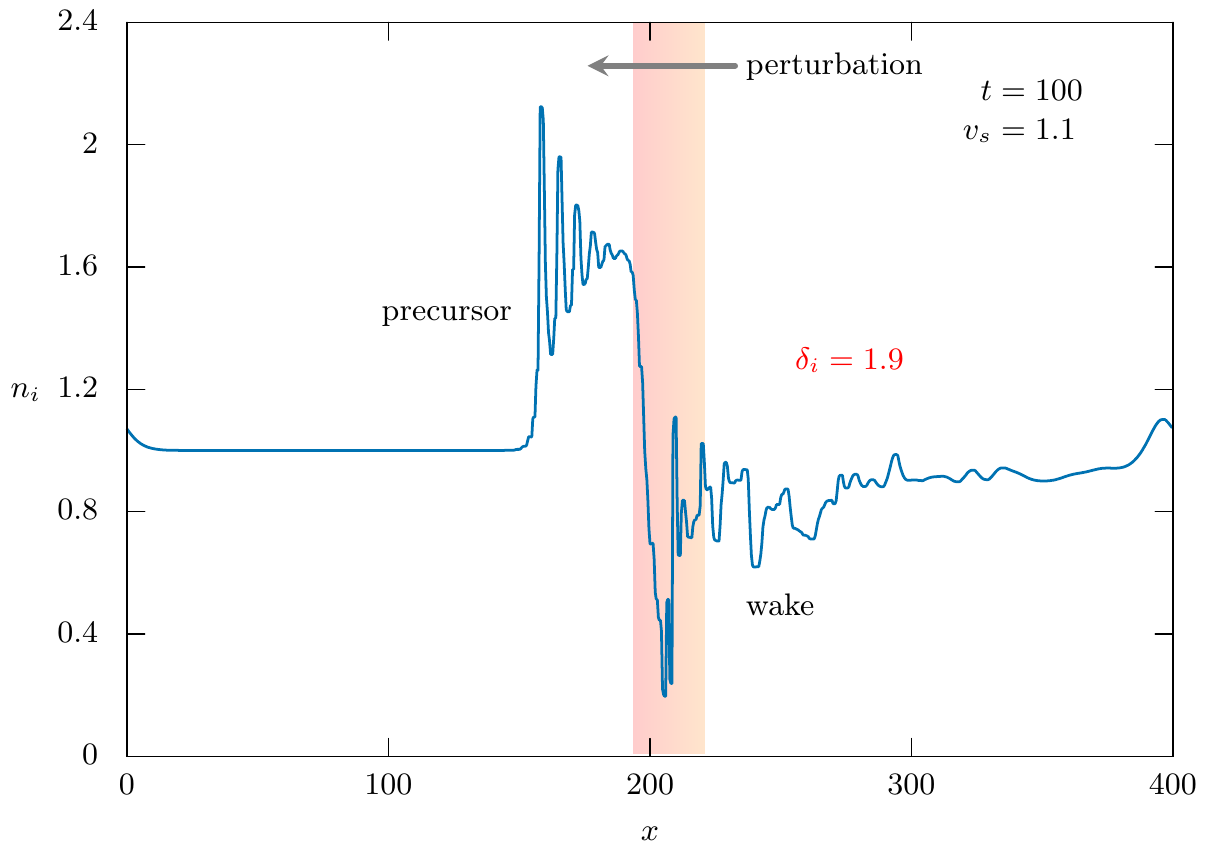}
\par\end{centering}
\caption{\label{fig:Top-panel-:}Top panel : Formation of dispersive shock
waves for high perturbation level as velocity increases. Bottom panel
: The same as above but with dust present. As we can see that presence
of dust suppresses formation of the shock front at low velocity $v_{s}=0.2$
(First figure of the bottom panel).}
\end{figure}

\subsubsection{Positive charge perturbation}

At low velocity (of the external perturbation or debris), these waves
propagates in either direction as multiple solitons as already seen
in the second panel of Fig.\ref{fig:Response-of-the}. Naturally the
precursors are compressed as compared to the wakes due to the moving
debris. However, as the speed approaches $v_{s}=1$, we see formation
of the so-called \emph{pinned} solitons, which are localized at the
perturbation site (and so the name pinned). In Fig.\ref{fig:The-precursor,-wakes,},
we have shown these two structure at two velocities $v_{s}=0.2$ and
$0.5$, after an evolution of $100$ plasma periods, with an without
dust particles. As mentioned in the previous subsection, the dust
particles effectively enhances the sound velocity thereby lowering
a perturbation to subsonic level which prevents the formation pinned
solitons. This can be clearly seen from Fig.\ref{fig:The-precursor,-wakes,}
where we have shown the formation of pinned solitons, precursors,
and wakes for a relatively low positive charge perturbation. One should
compare the 1st figure in the bottom panel with the 2nd figure of
the top panel, both of which are same except that in the bottom panel
dust particles are present which prevents the formation of pinned
solitons even at high velocity making the perturbation comparable
to low-velocity perturbation (1st figure of the top panel). In all
the cases and in the following, we have taken $\delta_{i}=1.9$, whenever
dust particles are present. In all the figures, we have shown the
site of the charge perturbation with a vertical stripe.

\begin{figure}[H]
\begin{centering}
~\hskip-24pt~\includegraphics[width=0.55\textwidth]{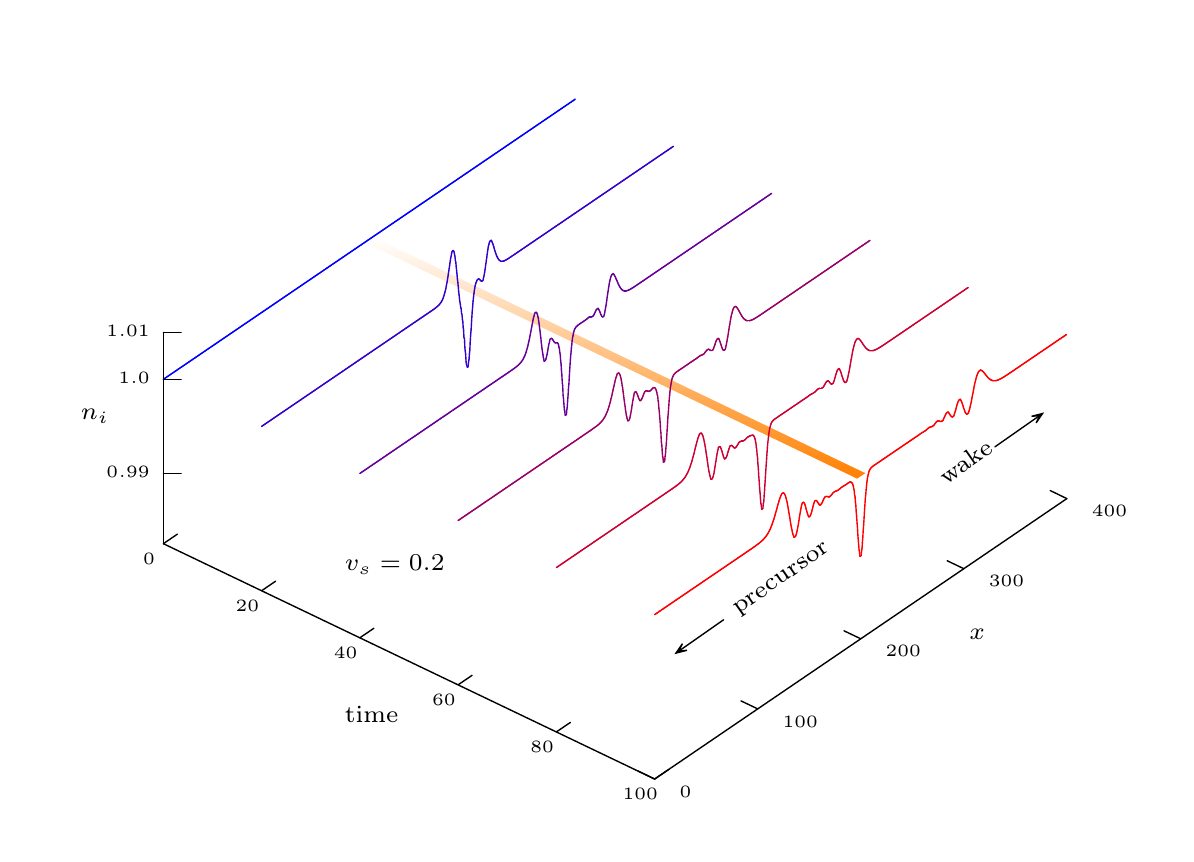}~\hskip-24pt~\includegraphics[width=0.55\textwidth]{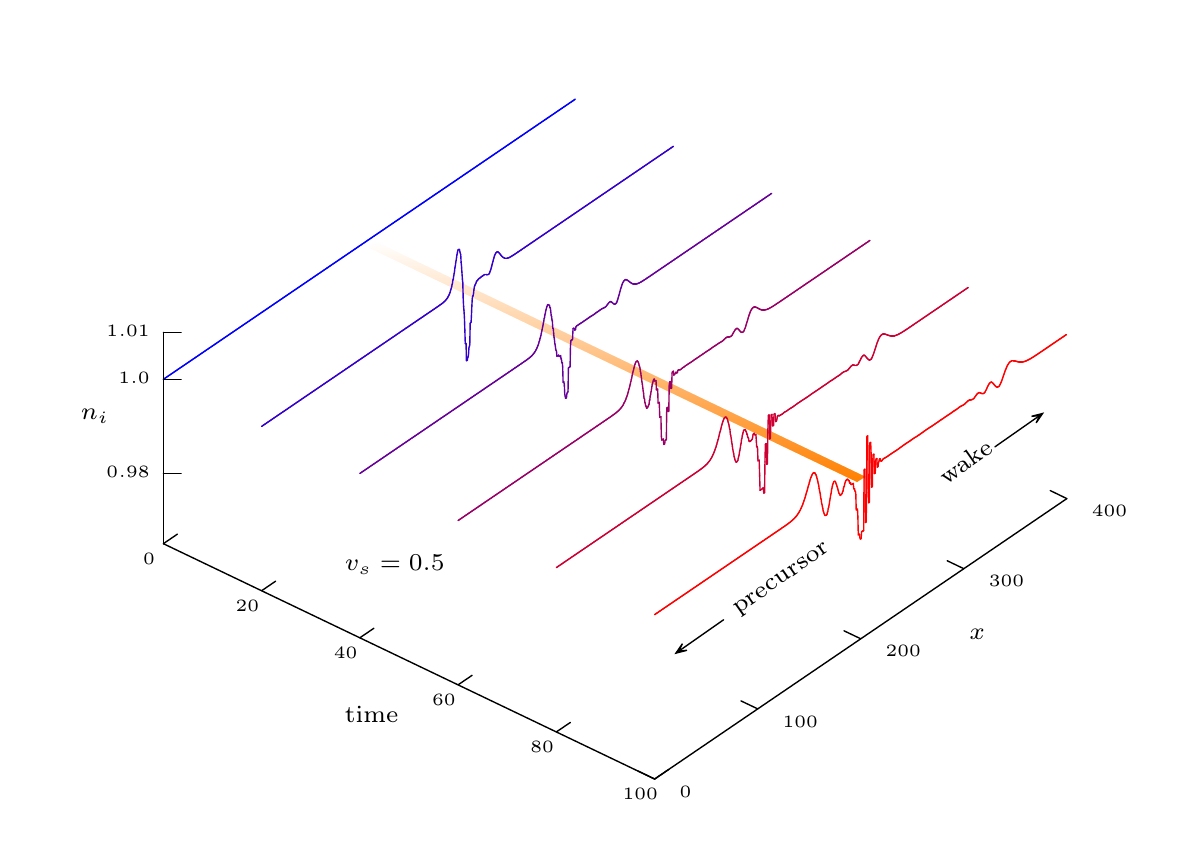}
\par\end{centering}
\begin{centering}
~\vskip-1.25in~\includegraphics[width=0.55\textwidth]{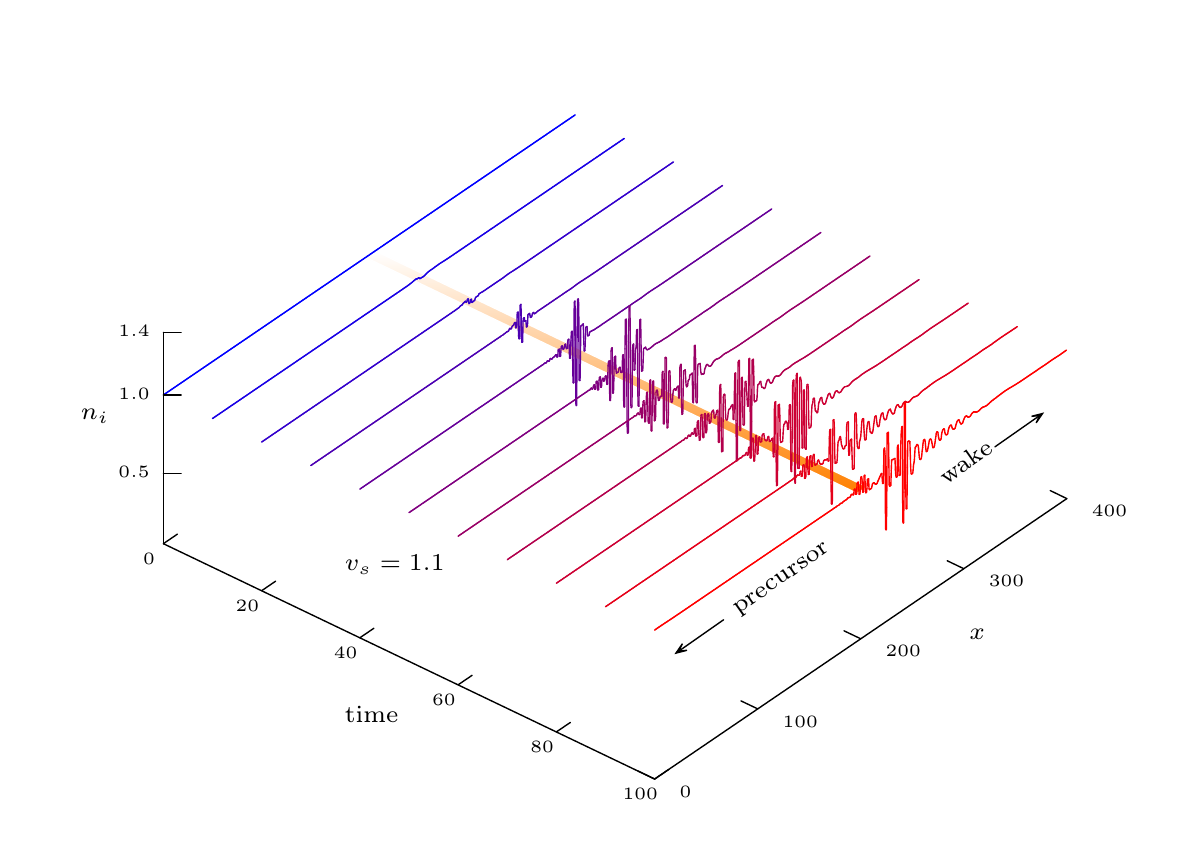}~\vskip-1in~
\par\end{centering}
\begin{centering}
~\hskip-24pt~\includegraphics[width=0.55\textwidth]{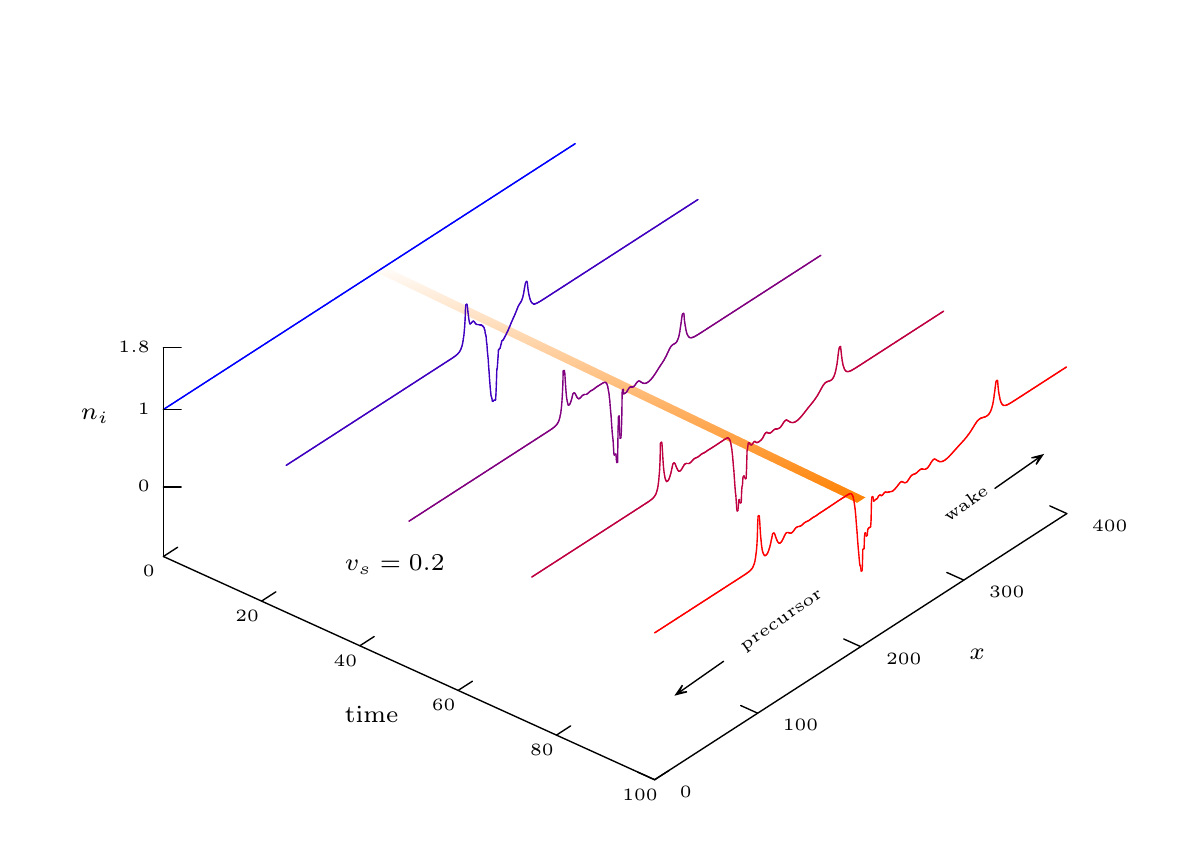}~\hskip-24pt~\includegraphics[width=0.55\textwidth]{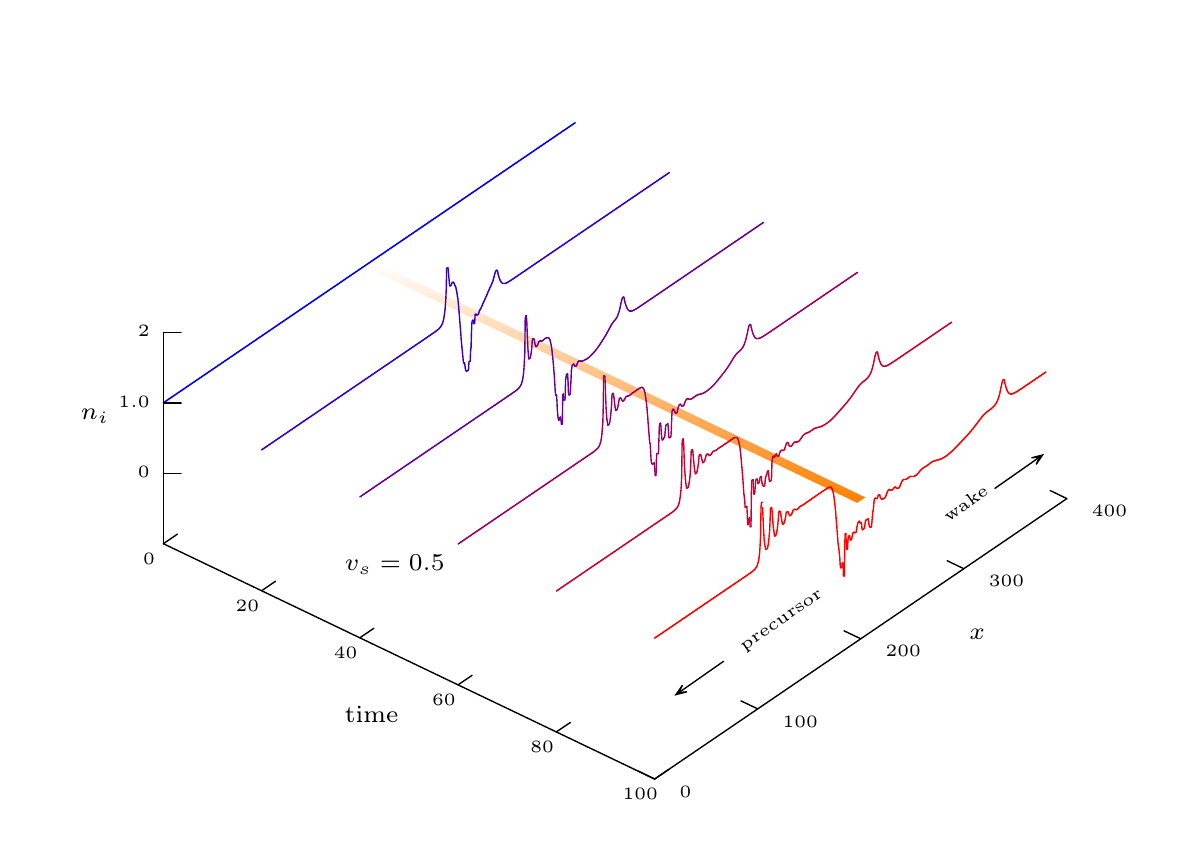}
\par\end{centering}
\begin{centering}
~\vskip-1.25in~\includegraphics[width=0.55\textwidth]{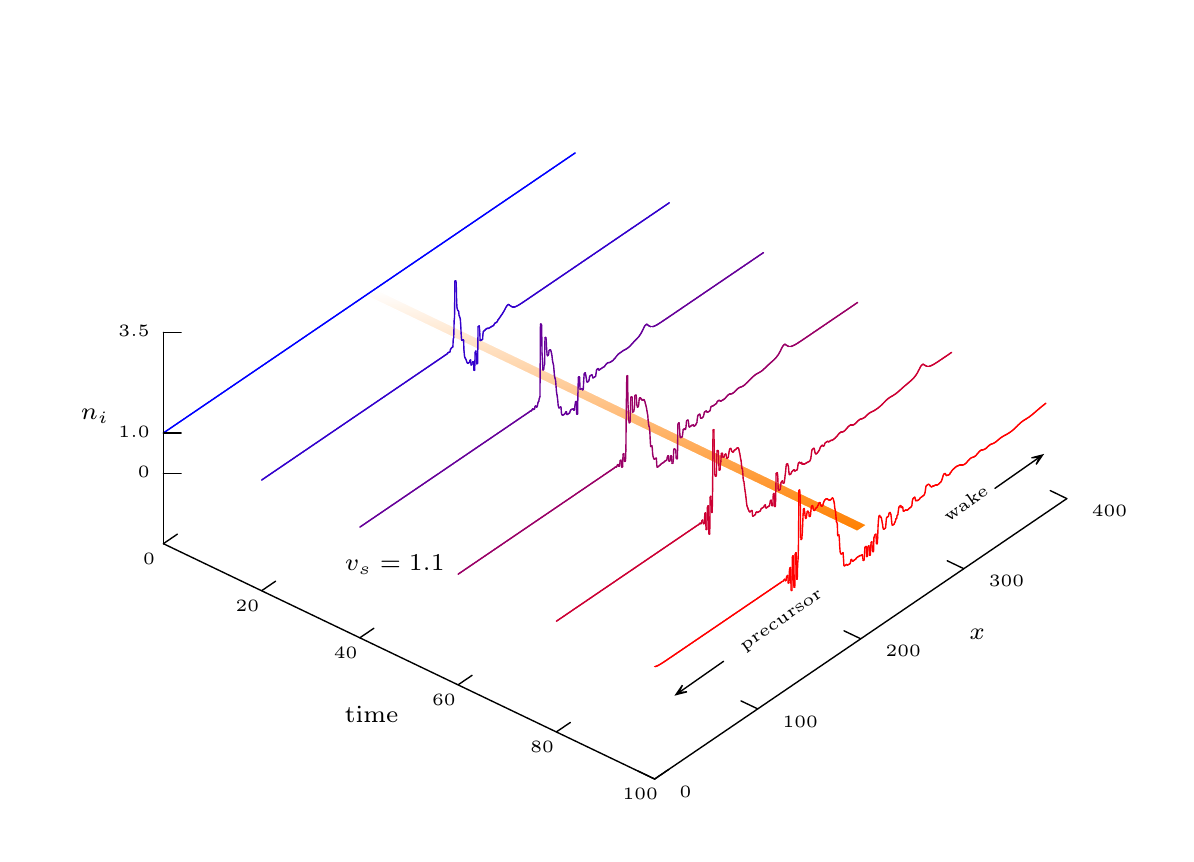}
\par\end{centering}
\caption{\label{fig:Full-time-evolution-of}Full time-evolution of a positive
charge perturbation without any dust effects. While top three figure
show the effect due to a relatively low perturbation level, the bottom
three figures show the response due to high perturbation. One can
clearly see the formation of the DSWs in the latter.
The orange stripe indicates the site of the perturbation.}
\end{figure}

\begin{figure}[t]
\begin{centering}
~\hskip-24pt~\includegraphics[width=0.55\textwidth]{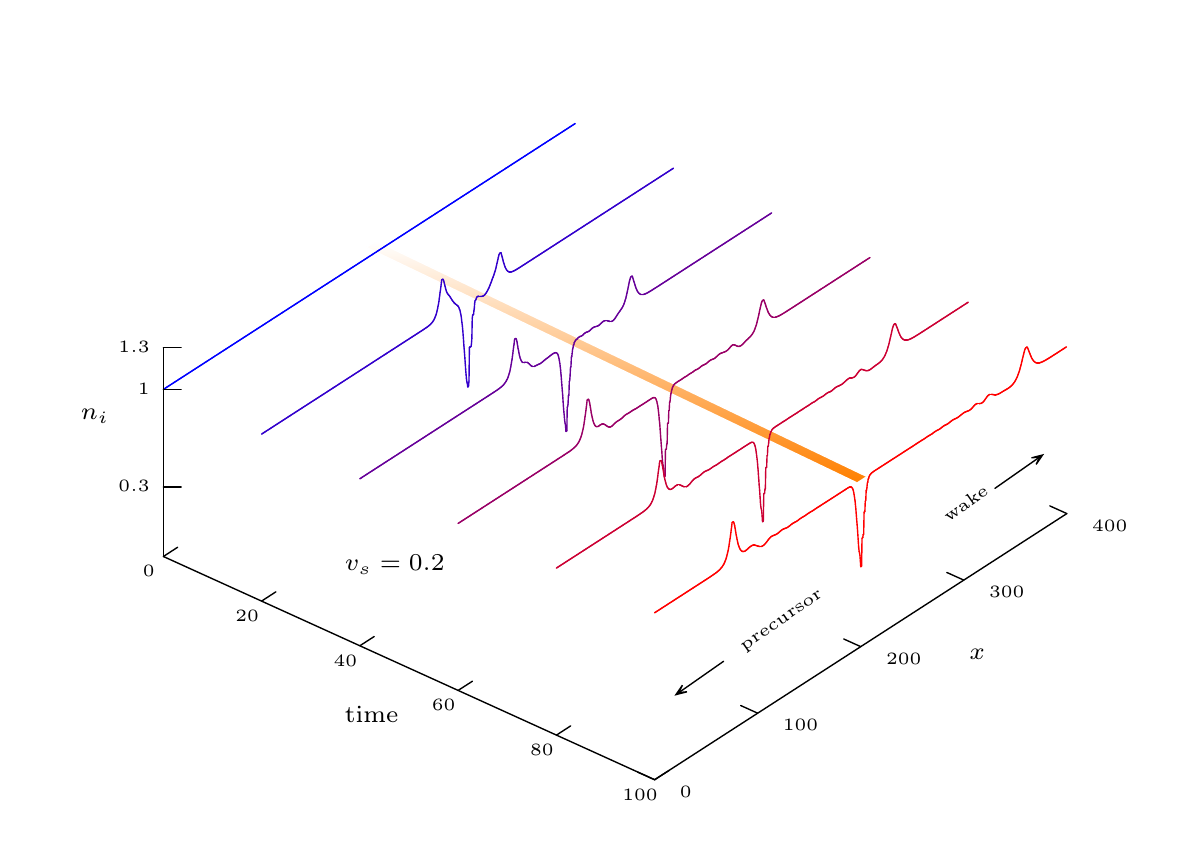}~\hskip-24pt~\includegraphics[width=0.55\textwidth]{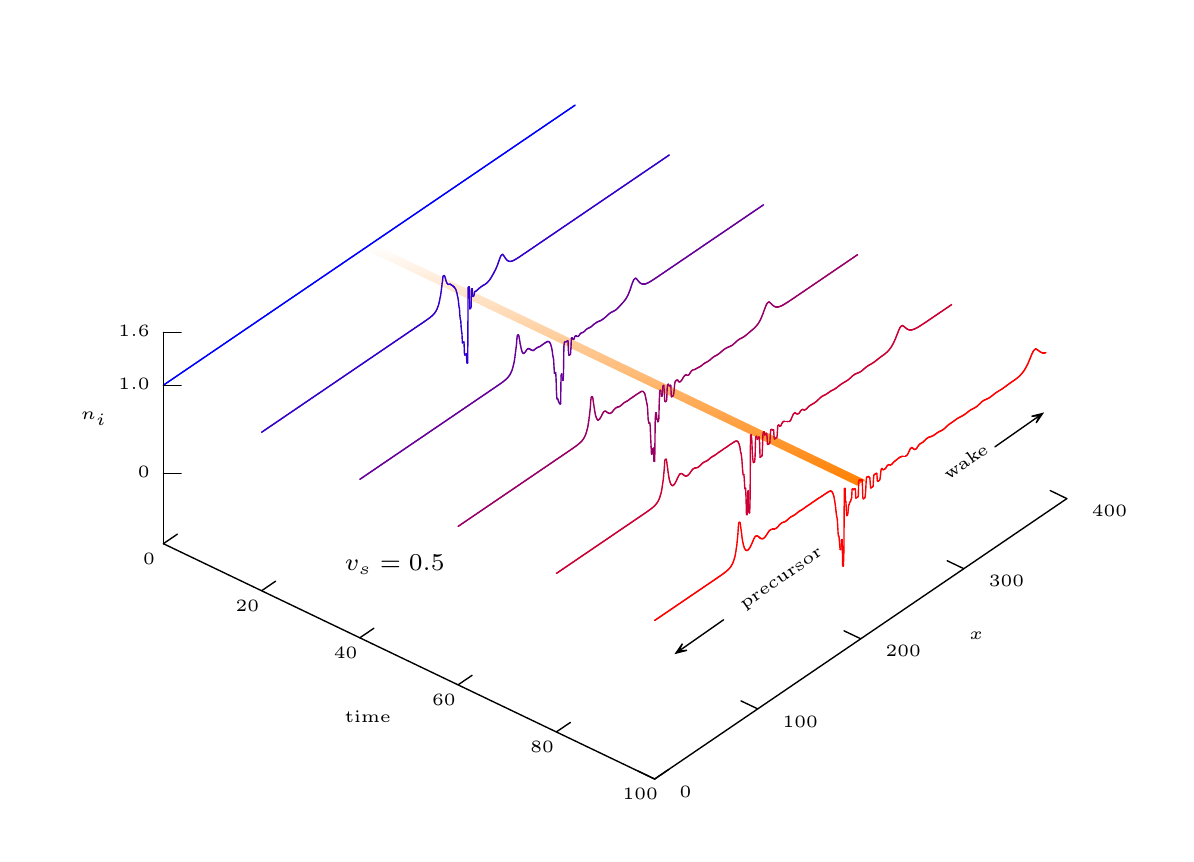}
\par\end{centering}
\begin{centering}
~\vskip-1.25in~\includegraphics[width=0.55\textwidth]{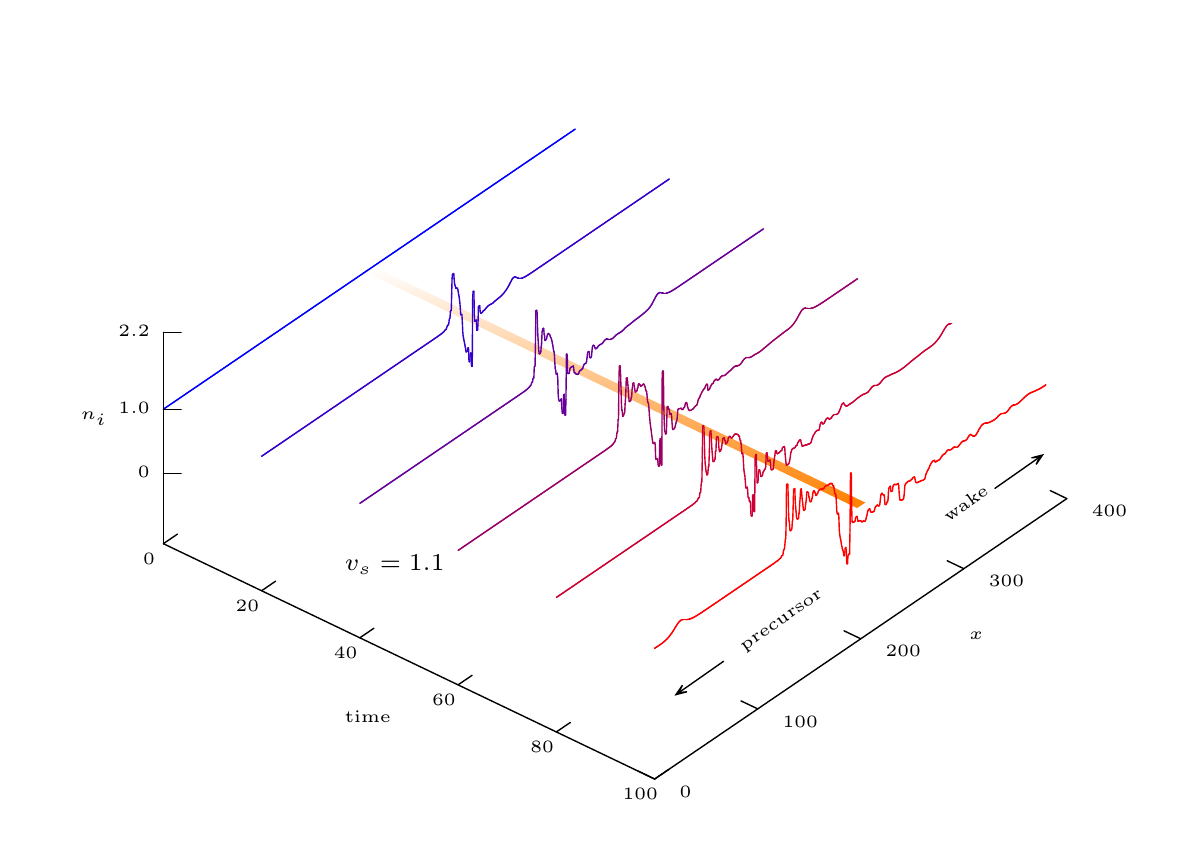}
\par\end{centering}
\caption{\label{fig:Response-to-a}Response to a high level of positive perturbation
in presence of dust $(\delta_{i}=1.9)$. One can clearly see the suppression
of formation of DSWs  at low velocities. At supersonic velocity
$(v_{s}=1.1)$ however, the suppression is not effective.}
\end{figure}

With high perturbation, we can see the formation of the DSWs in Fig.\ref{fig:Top-panel-:}.
The effect of dust basically suppresses the formation of DSWs at low
velocity $v_{s}=0.2$ at which DSW starts forming. In Fig.\ref{fig:Full-time-evolution-of},
we have shown the full time-evolution of the response of the plasma
to a positive charge perturbation (low and high) with any dust effects
and in Fig.\ref{fig:Response-to-a}, we show the time-evolution for
a high positive perturbation in presence of dust, where we can see
the suppression of formation DSWs due to the dusts.

\subsection{A theoretical approach}

A well-studied equation relevant in case of DSW is the nonlinear Schr\"odinger
equation, which we derive in this section for our plasma model. We
however leave out the dust-charge fluctuation dynamics as the dust-charge
fluctuation will lead to a dissipation term making it difficult to
isolate the DSWs.

The system of equations to be solved are
\begin{eqnarray}
\frac{\partial n_{i}}{\partial t}+\frac{\partial}{\partial x}(n_{i}v_{i}) & = & 0,\label{eq:cont-2}\\
n_{i}\frac{dv_{i}}{dt} & = & -\sigma\frac{\partial n_{i}}{\partial x}-n_{i}\frac{\partial\phi}{\partial x},\label{eq:mom-2}\\
\frac{\partial^{2}\phi}{\partial x^{2}} & = & n_{e}-\delta_{i}n_{i}+\delta_{d}z_{d}+\rho(x-v_{s}t),\label{eq:pois-2}
\end{eqnarray}
where the variables are normalised as before with $\rho$ as the normalised
external charge perturbation. At this time, we neglect the dust charge
fluctuation. Besides, as $\sigma\ll1$ we can neglect the ion thermal
pressure term in Eq.(\ref{eq:mom-2}) for the sake of simplicity.
In this limit we need to consider the current balance equation $I_{i}+I_{e}=0$,
where $I_{i,e}$ being the currents to the dust particles due to the
ions and electrons given by Eqs.(\ref{eq:ii},\ref{eq:ie}). So, the
charging equation Eq.(\ref{eq:charging}) becomes
\begin{equation}
\delta_{i}\delta_{m}\sigma^{1/2}\left(1-\frac{\varphi_{d}}{\sigma}\right)-\exp(\phi+\varphi_{d})=0,
\end{equation}
where we have replaced the ion density by its equilibrium value (unity)
for the sake of simplicity. This can be justified considering the
fact that electrons are almost inertia-less in comparison to the ions
and most of the current to the dust particles contributed by the electrons.
This equation can be solved in terms of Lambert $W$ function
\begin{equation}
z_{d}=\frac{W(\vartheta e^{\phi})-\sigma}{W(\vartheta)-\sigma},\label{eq:zd}
\end{equation}
where $\vartheta=\sqrt{\sigma}e^{\sigma}/(\delta_{i}\delta_{m})$. 

In order to investigate the occurrence of dispersive shock, we use
the reductive perturbation method to derive the nonlinear Schr\"odinger
equation (NLSE) for our model. We use the stretched variables as
\begin{equation}
\xi=\varepsilon(x-ut),\quad\tau=\varepsilon^{2}t,\label{eq:coord}
\end{equation}
corresponding to the space and time variables, where $\varepsilon$
is the small expansion parameter and $u$ is the phase velocity of
the wave. We expand the dependent variables as
\begin{equation}
F(x,t;\xi,\tau)=F_{0}+\sum_{j=1}^{J}\varepsilon^{j}\sum_{l=-L}^{L}f_{j,l}(\xi,\tau)e^{il(kx-\omega t)},\label{eq:expansion}
\end{equation}
where
\begin{eqnarray}
F & = & (n_{i}v_{i},\phi)',\\
f & = & (n,v,\phi)',
\end{eqnarray}
with $n_{i0}=1$ and $v_{i0}=\phi_{0}=0$. The external charge perturbation
term $\rho(x-v_{s}t)$ is assumed to be a constant and second order
term
\begin{equation}
\rho\equiv\varepsilon^{2}s.
\end{equation}
The space and time derivative can be mapped to the new set of variables
as
\begin{eqnarray}
\frac{\partial}{\partial t} & \to & \frac{\partial}{\partial t}+\varepsilon^{2}\frac{\partial}{\partial\tau}-\varepsilon u\frac{\partial}{\partial\xi},\\
\frac{\partial}{\partial x} & \to & \frac{\partial}{\partial x}+\varepsilon\frac{\partial}{\partial\xi}.
\end{eqnarray}
We note that the electron density and dust-charge number can be expanded
in terms of plasma potential $\phi$ as 
\begin{figure}[t]
\begin{centering}
\includegraphics[width=0.5\textwidth]{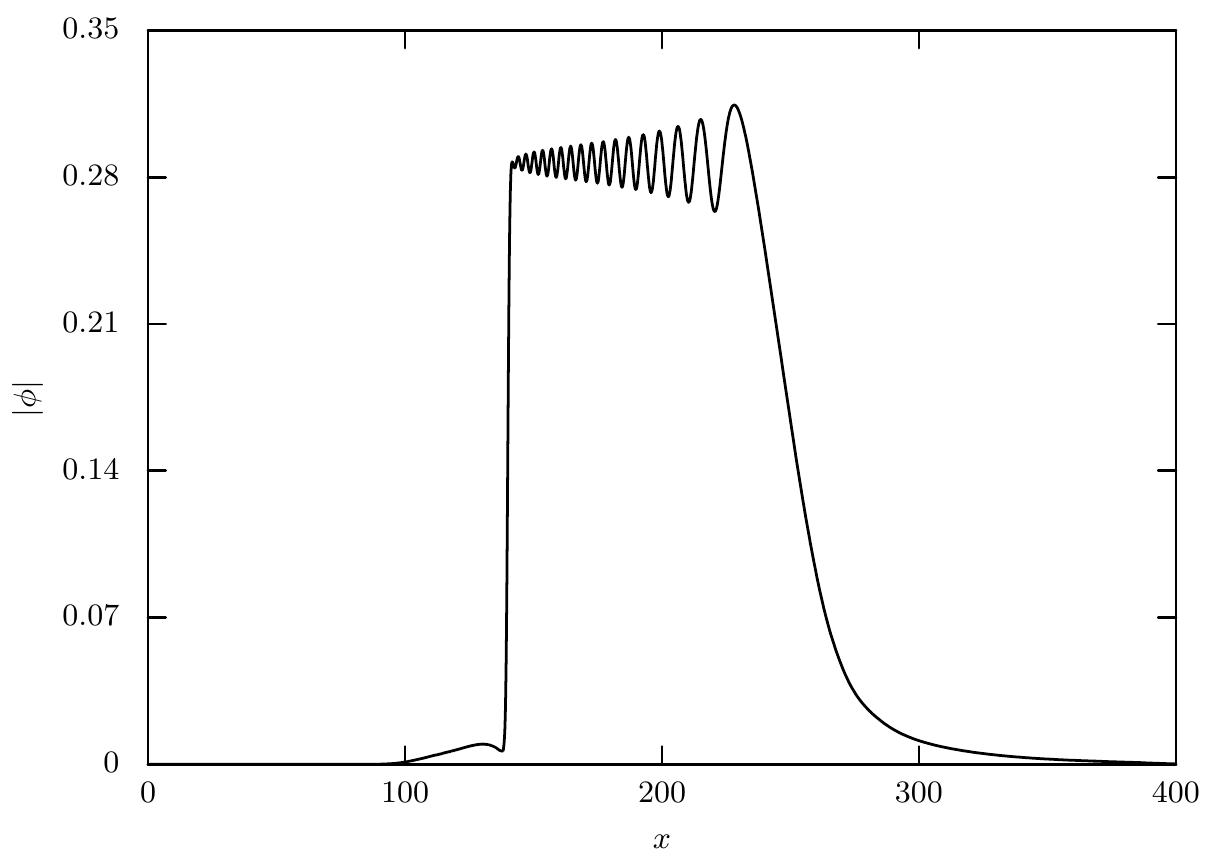}\hfill{}\includegraphics[width=0.5\textwidth]{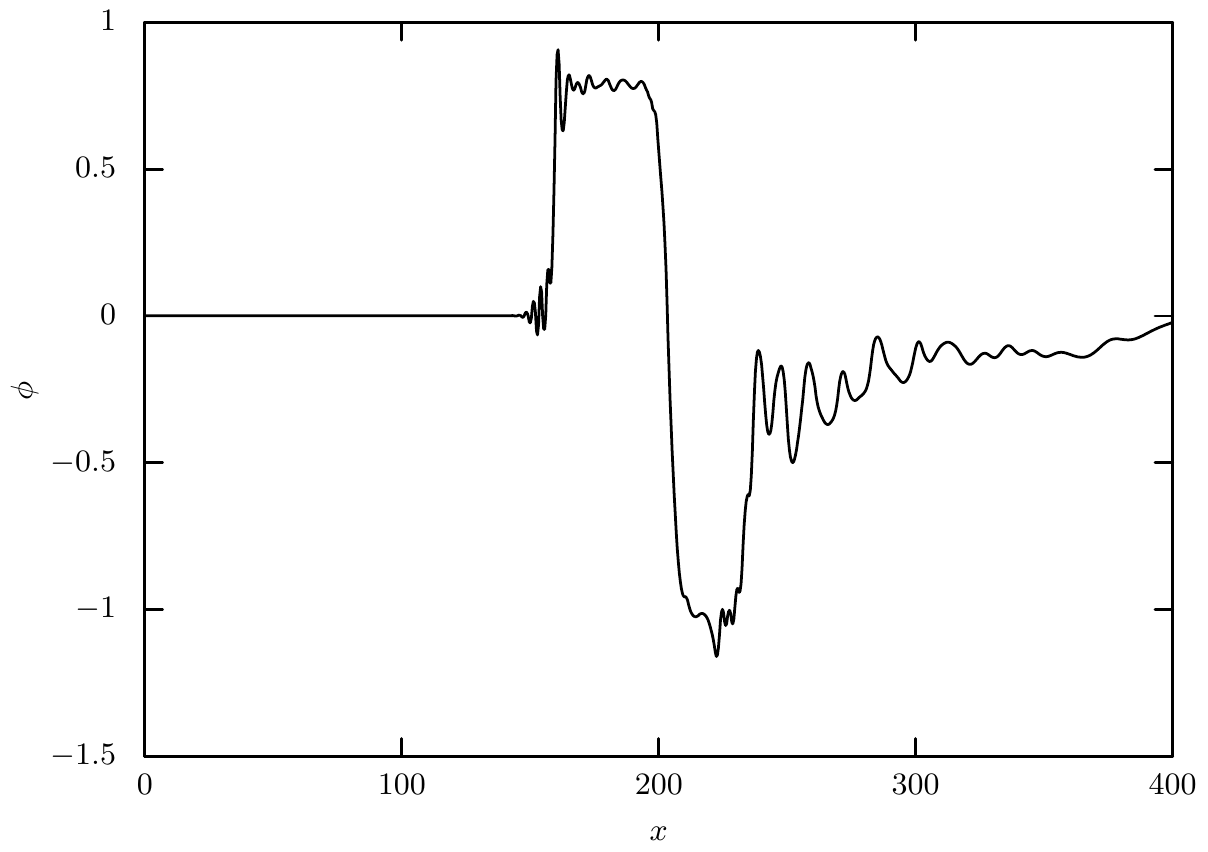}
\par\end{centering}
\caption{\label{fig:Comparison-of-DSWs}Comparison of DSWs obtained from NLSE
and FCT simulation for $v_{s}=1.1$.}
\end{figure}
\begin{align}
n_{e} & \simeq1+\phi+\frac{1}{2}\phi^{2}+\frac{1}{6}\phi^{3}+\cdots,\\
z_{d} & \simeq1+c_{1}\phi+c_{2}\phi^{2}+c_{3}\phi^{3}+\cdots.\label{eq:exp}
\end{align}
The first of the above is due to the Boltzmannian nature of the electrons
and second expression is derived by expanding Eq.(\ref{eq:zd}) about
$\phi=0$, where
\begin{eqnarray}
c_{1} & = & \frac{W_{1}}{(1+W)},\\
c_{2} & = & \frac{W_{1}}{2(1+W)^{3}},\\
c_{3} & = & \frac{W_{1}(2W-1)}{6(1+W)^{5}},\label{eq:c3}
\end{eqnarray}
with $W_{1}=W/(\sigma-W)$ and $W\equiv W(\vartheta)$. Numerically,
the expansion (\ref{eq:exp}) is remarkably accurate up to a fairly
large $\phi$. For example for $\phi$ as high as unity, the difference
between Eqs.(\ref{eq:zd},\ref{eq:exp}) is just about $\sim0.02\%$
and we can safely replace $z_{d}$ by the above expansion in our calculations.

Following the standard procedure (see Appendix), we reduce all the
expressions to an NLSE in the first order variable $\phi_{1,1}$
\begin{equation}
i\frac{\partial\phi_{1,1}}{\partial\tau}+C_{1}\frac{\partial^{2}\phi_{1,1}}{\partial\xi^{2}}+C_{2}|\phi_{1,1}|^{2}\phi_{1,1}=C_{3}s\phi_{1,1},\label{eq:nlse}
\end{equation}
which is also known as Gross-Piteavskii equation (GPE), usually found
in the context of Bose-Einstein condensation. This equation also appears
and in case of dispersive hydrodynamic flow past an obstacle, known
as the \emph{piston} problem -- a mechanical equivalent of our problem.
We note that while the same problem can also be studied with the help
of the Korteweg-de-Vries (K-dV) equation (the so-called forced K-dV)
equation, we believe that the NLSE describes the situation more appropriately
as an NLSE accounts for the slow time and space modulation of a linear
wave through the variations in the medium itself and the nonlinear
effects, while a K-dV equation describes the soliton which is produced
due to the delicate balance between nonlinear effects and dispersion.
We further note that the NLSE can be derived from a K-dV equation
by considering the solution to be a wave packet -- thereby generalizing
the form of nonlinear waves supported in a particular situation. In
Fig.\ref{fig:Comparison-of-DSWs}, we show generation of DSW through
numerical solution of Eq.(\ref{eq:nlse}) in presence of only external
perturbation $(s\neq0)$ alongside the simulation results for $\phi$.
The NLSE has been solved using Mathematica with periodic boundary
conditions and initial conditions $\phi_{1,1}=0$. The functional
form for $s$ is same as in Eq.(\ref{eq:pulse}). We have used the
career wave number $k=1$ in the NLSE. Though we usually solve for
$n_{i}$ in our FCT simulation, for the sake of comparison, in Fig.\ref{fig:Comparison-of-DSWs},
we have shown the results for $\phi$, which closely resembles $n_{i}$
which is quite justified in the ion-acoustic regime. One can clearly
see the similarities between the theoretical result with that from
FCT simulation. In both cases, the debris velocity $v_{s}=1.1$ and
perturbation amplitude $\hat{\rho}=1$.

\subsubsection{Effect of dusts}

\begin{figure}[t]
\begin{centering}
\includegraphics[width=0.5\textwidth]{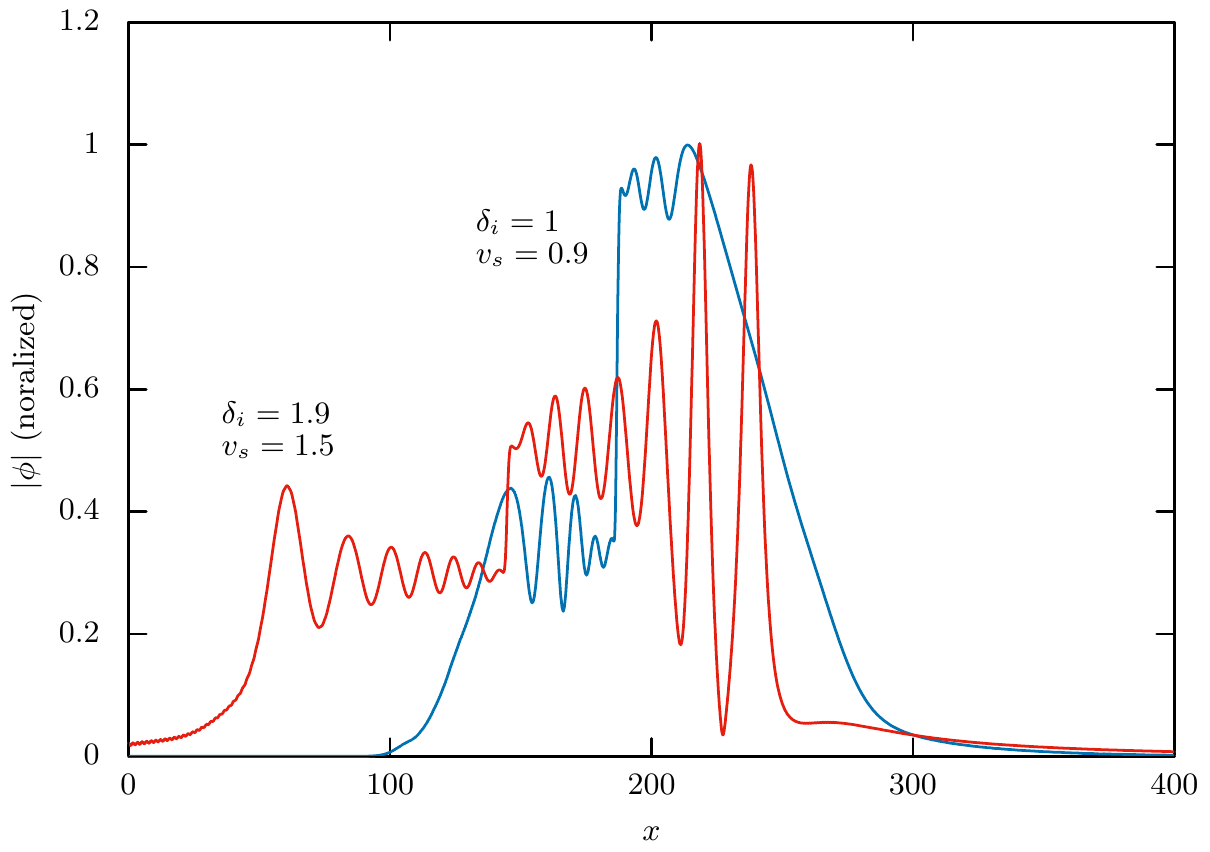}
\par\end{centering}
\caption{\label{fig:Effect-of-negatively}Effect of negatively charged dust
particles on formation of DSW in NLSE solutions. The blue colored
curve is for the case with no dust $(\delta_{i}=1)$, where a DSW
ie beginning to form at around $v_{s}\simeq0.9$. While the red colored
curve shows the same case with dust particles $(\delta_{i}=1.9)$,
where a DSW is starting to form \emph{only} at around $v_{s}\simeq1.5$.}
\end{figure}

As we have seen from our simulation results, the primary effect of
negatively charged dust particles is to increase the effective ion-acoustic
velocity via Eq.(\ref{eq:ia}). The same effect is visible in the
NLSE solutions as well. In Fig.\ref{fig:Effect-of-negatively}, we
have compared the the cases for formation of DSW through solutions
of the NLSE {[}Eq.(\ref{eq:nlse}){]} when there is no dust particles
$(\delta_{i}=1)$ and in presence of dust particles $(\delta_{i}=1.9)$.
All other parameters are kept same. As we can see that while the DSW
begins to form at around $v_{s}\simeq0.9$ in case of the former,
they start to form \emph{only} at about $v_{s}\simeq1.5$ when dust
particles are present. Theoretically from Eq.(\ref{eq:ia}), we have
an enhancement factor of $\sim1.38$ of the ion-acoustic velocity
for $\delta_{i}=1.9$. This tells us that the DSW might form \emph{only}
at a debris velocity $v_{s}\gtrsim1.24$ when $\delta_{i}=1.9$, assuming
that DSW forms at about $v_{s}\simeq0.9$ when $\delta_{i}=1$. This
very much agree with our solutions.

\begin{figure}[t]
\begin{centering}
\includegraphics[width=1\textwidth]{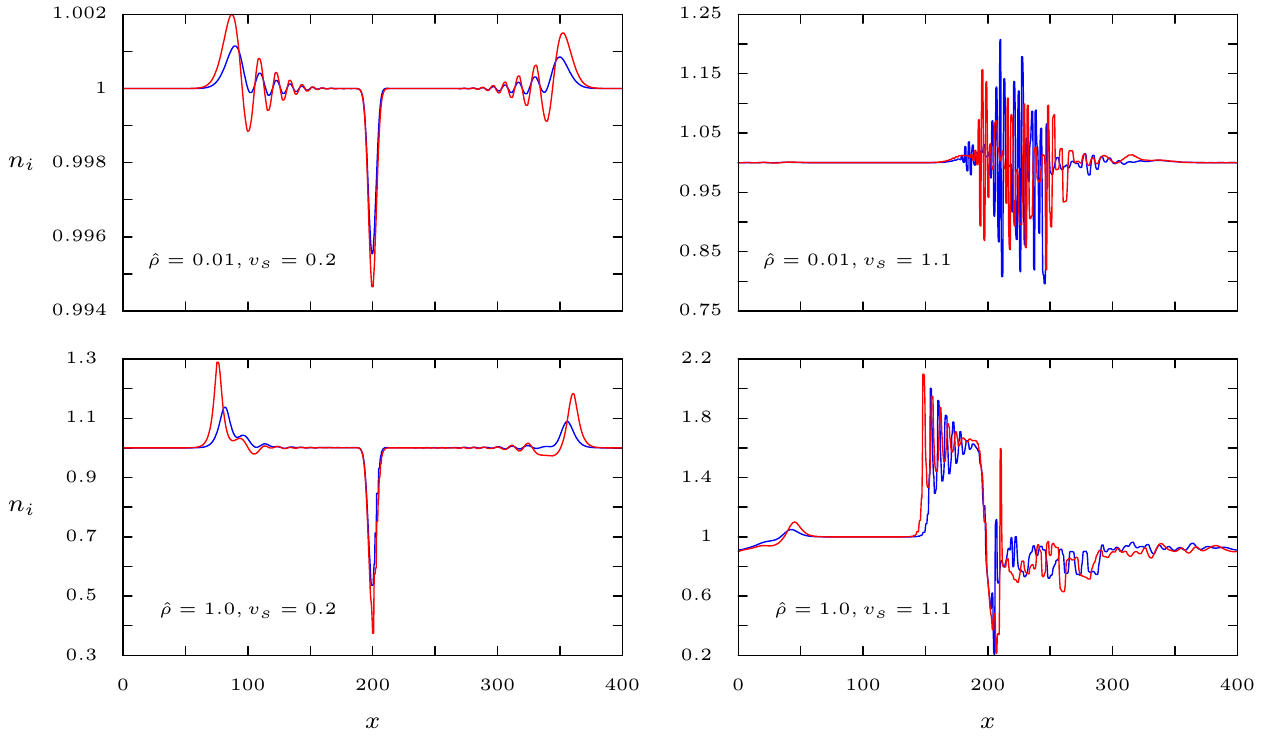}
\par\end{centering}
\caption{\label{fig:The-of-dust-charge}The effect of dust-charge fluctuation
on formation of DSW. The top panel shows the evolution at low perturbation
(with low and high debris velocities). The bottom panel shows the
same at high perturbation (with low and high debris velocities).}
\end{figure}

In Fig.\ref{fig:The-of-dust-charge}, we have shown the effect of
dust-charge fluctuation on formation of DSW. As we have seen before,
the DSW forms only at high debris velocities. In all cases, the effect
of dust-charge fluctuation is \emph{only }minimal causing a slight
damping of the oscillations, which is what is expected.

\section{Negative charge perturbation}

\begin{figure}[t]
\begin{centering}
\includegraphics[width=1\textwidth]{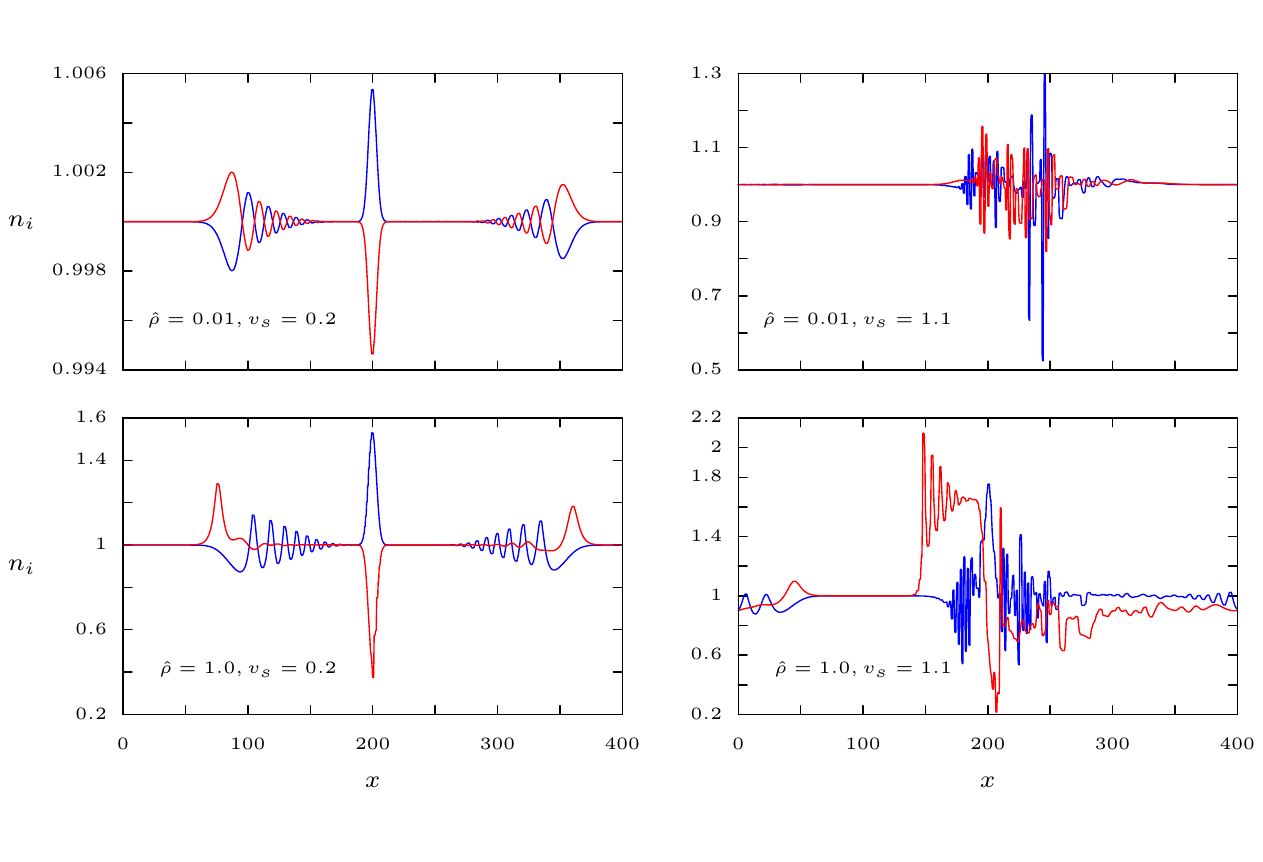}
\par\end{centering}
\caption{\label{fig:Response-to-negative}Response to a negative charge perturbation,
which shows complete suppression of DSWs. The top panel shows the
evolution at low perturbation (with low and high debris velocities).
The bottom panel shows the same at high perturbation (with low and
high debris velocities).}
\end{figure}

We note that, usually in a plasma, the nature of charge perturbation
should not matter as it propagates to all physical variables as time
progresses. However, when we have an external charge perturbation
such as caused by the presence of debris, which acts as constant source
of perturbation, the response of the plasma can be quite different.
In the ion-acoustic regime we can understand this from the fact that
an ion-acoustic wave, especially in the low frequency regime, propagates
primarily due to the inertia of the ions tugging the electrons along.
Naturally, any response of the plasma to an external perturbation
depends on how do the ions respond. As we can see from Fig.\ref{fig:Response-to-negative},
while for low debris velocity, the response of the plasma to positive
and negative charge perturbations are almost out of phase, as expected.
However, when the debris velocity is high, especially in the large
perturbation regime, the formation of so-called DSWs are suppressed.

\section{Summary and conclusions}

We investigated the dynamics of ion-acoustic waves in a dusty plasma
for an unmodulated external moving charge perturbation using FCT simulation,
for three different environments -- (\emph{i}) in absence of dust
grains, (\emph{ii}) in presence of dust grains, and (\emph{iii}) in
presence dust-charge fluctuations. The morphology as well the response
of the plasma at the site of perturbation is found to be different
for different nature of the charge perturbations. As a matter of fact,
the site of the perturbation behaves like a non-quasi-neutral plasma
for the case of constant charge perturbation, other than that, the
rest of the plasma remains quasi-neutral, similar to that of a density
perturbed plasma. We have observed the plasma in presence of both
positive and negative charge perturbations in the form of an external
gaussian pulse for two different strength -- one large and other
relatively small charge perturbation. The results of our simulation
are quite intriguing.

Our first significant observation from the simulation is that an unmodulated
moving external charge perturbation is self-sufficient to excite nonlinear
IA oscillations and both precursor and wake waves in the plasma. Earlier
theories and observations \citep{Jaiswal2016pre,Arora2019,Tiwari2016}
suggest that in without a modulating function associated with the
source function can only give rise to wake fields at subsonic and
precursor and pinned solitons at supersonic velocities. Further it
was suggested that only a modulating perturbation, be it density perturbation,
velocity perturbation or charge perturbation \citep{Chakraborty2022}
can lead to an electrostatic IA shock. However, from our simulation
of an unmodulated charge perturbation, we can see the formation of
both precursors and wakes at subsonic velocity as well as in supersonic
velocity for a small charge perturbation and if the charge perturbation
becomes sufficiently large, then formation of the precursor solitons
as well as the dispersive shocks are observed.

A number of authors \citep{Sen2015,tiwari20162} have already predicted
the formation of precursor solitons which are pinned to the external
moving source. We on the other hand observe the formation of pinned
envelope solitons in the wake of the moving charge perturbation for
a relatively small strength of the perturbation with supersonic speed.
At a low velocity (subsonic flow) only precursor and wake structures
are excited and they move away from the site of the perturbation at
a speed equal to that of the phase velocity of the plasma ($>$ source
velocity) when it undergoes IA oscillations. As we increase the source
velocity and it crosses the critical sonic velocity, the wakes leave
the source site with a velocity relatively higher than that of the
precursors, which become compressed with increase in their amplitude
and move with a lower relative velocity than that of the velocity
of wakes but still remains ahead of the source. Interestingly, in
this supersonic region, the source structure also responds to the
IA oscillations and as time evolves the hind-side of the source breaks
down to start forming envelope soliton structure and at very high
source velocity the fully formed envelope solitons are excited which
remains `pinned\textquoteright{} to the site of perturbation. This
behaviour is consistent with perturbation of small strength of any
nature be it positive or negative, even though their initial form
is different and out of phase. 

For a perturbation with sufficiently large strength of the external
charge source, the precursor waves transform into a dispersive shock
wave with its characteristic oscillatory trail and turbulence behind
the source site. Similar kind of structures are already observed theoretically
with modulating external source \citep{Chakraborty2022}, with molecular
dynamic simulation \citep{Tiwari2016} and also observed experimentally
\citep{Jaiswal2016,Jaiswal2016pre} for dust acoustic waves in a flowing
dusty plasma. But our simulation shows that kinds of electrostatic
ion-acoustic DSWs are only formed when the unmodulated external perturbation
happens to be positively charged for a considerably large strength.
Unlike precursor solitons, the DSWs starts forming even at a low velocity
and as we gradually increase the source velocity, they become more
compressed and larger in amplitude with increased turbulence in the
wake region. In fact, for a considerably high velocity, they become
so compressed that small amplitude envelope structures are seen to
be forming even ahead of the DSWs. However, no formation of envelope
solitons is seen neither at the perturbation site nor at the wake
field region in contrast to that of the small strength perturbation.
Interestingly for negative nature of the source charge, no formation
of such DSWs is observed even for a large strength of perturbation
in supersonic velocity. Only envelope solitons in the precursor and
wake region are formed as expected. Formation of the DSWs is suppressed
when the external charge perturbation is negative since positive electrons
are accumulated at the perturbation site making the precursors more
compressed with the increasing source velocity and making them pinned
ahead of the source. 

The amplitudes of the envelope solitons and the DSWs are susceptible
to the nonlinear IA oscillations and can vary with it as time evolves.
Dust effects are very distinct in terms of both amplitudes and modification
of the effective transcritical velocity. Presence of dust particles
reduces the nonlinear structure amplitudes as the \emph{effective}
ion-sound speed gets increased proportionally by a factor depending
on the concentration of the dust present, which remains consistent
with the theory\citep{Shukla_Mamun_2015}. Due to which, in contrast
to the charge disturbance moving through a dustless media, even at
a moderately high velocity, the ion density does not show formation
of envelope solitons. Similarly, dispersive IA shocks are suppressed
by the presence of dust particles at a low velocity unlike $e$-$i$
plasma, where shocks are formed even at low velocities. But for considerably
high velocity, the modification of the ion-sound speed becomes insignificant
and fully formed envelope solitons and IA DSWs are observed. Effects
of dust-charge fluctuation is not very significant except causing
a decrease in amplitudes of the excitations, which was expected since
IA oscillations get dampened under the influence of dust-charge fluctuations.
The NLSE reduced for our plasma model also validates our observations.
The NLSE solutions show the formation of DSWs with typical oscillatory
trails, as well as dust effects such as ion-sound speed enhancement. 

To conclude, we have demonstrated how pinned envelope solitons and
precursor DSWs are excited as a response of a dusty plasma to an external
unmodulated moving charge perturbation, by doing a detailed parametric
study in the nonlinear regime. The circumstances surrounding their
occurrence is found to be directly related to the nature ($+{\rm ve}$
or $-{\rm ve}$) and strength of the external perturbation applied
($\hat{\rho}$) and also depend upon the environment of the media
through which it propagates. Theoretical estimates using the NLSE
approach also verifies the same dynamical behaviour. The results from
this study may have potential applications in relevant space or atmospheric
plasma technologies. We also consider our study to be experimentally
reproducible as the parameter region we are studying is well in accordance
with that of laboratory environments. 

\section*{Acknowledgements}

One of the authors H. Sarkar gratefully acknowledges the Junior Research
Fellowship (JRF) received from CSIR-HRDG, New Delhi, India {[}File
No: 09/059(0074)/2021-EMR-I{]}.

\section*{Appendix}

\subsection{Derivation of the NLSE}

The relevant equations to be reduced to NLSE are Eqs.(\ref{eq:cont-2}-\ref{eq:pois-2}).
We use the expansion and other related quantities as defined in Eqs.(\ref{eq:coord}-\ref{eq:c3})
and consider various orders in the expansion parameter $\varepsilon$,
as follows.

At the outset, we note that the following conditions can be proved
to be true
\begin{equation}
f_{j,l}=0,\left\} \quad\begin{array}{rcl}
j & > & 3\\
|l| & > & j\\
j & = & 1,l=0\\
j & = & 3,|l|>1
\end{array}\right.
\end{equation}
First, we consider the terms arising at the order $(j=1,l=1)$ from
Eqs.(\ref{eq:cont-2},\ref{eq:mom-2}), which provides the first order
density and velocity,
\begin{equation}
n_{1,1}=\frac{k^{2}}{\omega^{2}}\phi_{1,1},\quad v_{1,1}=\frac{k}{\omega}\phi_{1,1}.
\end{equation}
Inserting the above expressions into Eq.(\ref{eq:pois-2}) provides
the linear dispersion relation (or the first compatibility condition),
\begin{equation}
\omega=k\left(\frac{\delta_{i}}{B}\right)^{1/2},
\end{equation}
where $B=k^{2}+A$ and $A=1+c_{1}\delta_{d}$. In the limit of zero
dust particles, $\delta_{i}\to1,\delta_{d}\to0$ and the relation
reduces to the familiar ion-acoustic dispersion relation
\begin{equation}
\omega=\frac{k}{\sqrt{1+k^{2}}}.
\end{equation}

Next, we consider the terms at the order $(j=2,l=1)$ from Eqs.(\ref{eq:cont-2}-\ref{eq:pois-2}),
which yield the next order quantities,
\begin{eqnarray}
n_{2,1} & = & -2i\frac{k}{\omega^{3}}(\omega-ku)\,\partial_{\xi}\phi_{1,1},\\
v_{2,1} & = & -\frac{i}{\omega^{2}}(\omega-ku)\,\partial_{\xi}\phi_{1,1}.
\end{eqnarray}
We should note here that unless the condition $\phi_{2,1}=0$ is satisfied,
the above second order quantities become indeterminate. This also
results in the second compatibility condition in terms of the group
velocity of the wave
\begin{equation}
u=A\frac{\delta_{i}^{1/2}}{B^{3/2}},
\end{equation}
which again reduces to the usual ion-acoustic velocity $u=k/\sqrt{1+k^{2}}$
in the limit of zero dust.

The next order to be considered is $(j=2,l=2)$, which provides the
$f_{2,2}$ quantities,
\begin{eqnarray}
\phi_{2,2} & = & \frac{1}{C}\left(\omega^{2}D-3k^{4}\delta_{i}\right)\phi_{1,1}^{2},\\
n_{2,2} & = & \frac{1}{C}\left[D-3k^{2}(A+4k^{2})\right]\phi_{1,1}^{2},\\
v_{2,2} & = & \frac{k}{\omega C}\left[\omega^{2}D-k^{2}\left\{ A\omega^{2}+2k^{2}\left(\delta_{i}+2\omega^{2}\right)\right\} \right]\phi_{1,1}^{2},
\end{eqnarray}
where $C=2\omega^{2}\left[k^{2}\delta_{i}-\omega^{2}(A+4k^{2})\right]$
and $D=\omega^{2}(1+2\delta_{d}c_{2})$.

The next two orders are $(j=3,l=0)$ and $(j=2,l=0)$ which provides
the $f_{2,0}$ quantities,
\begin{eqnarray}
\phi_{2,0} & = & \frac{\omega u^{2}D-k^{2}\delta_{i}(\omega+2ku)}{\omega^{3}(\delta_{i}-Au^{2})}\,\left|\phi_{1,1}\right|^{2},\\
n_{2,0} & = & \frac{\omega D-Ak^{2}(\omega+2ku)}{\omega^{3}(\delta_{i}-Au^{2})}\,\left|\phi_{1,1}\right|^{2},\\
u_{2,0} & = & -\frac{\omega uD-k^{2}(2k\delta_{i}+\omega ku)}{\omega^{3}(\delta_{i}-Au^{2})}\,\left|\phi_{1,1}\right|^{2}.
\end{eqnarray}
The next order terms are $(j=3,l=1)$ which provides the expressions
for $f_{3,1}$ terms. However, as before unless $\phi_{3,1}=0$, the
solutions become indeterminate. So, one needs to re-calculate the
expressions for $(n,u)_{3,1}$ from Eqs.(\ref{eq:cont-2},\ref{eq:mom-2}).
By substituting these expressions in Eq.(\ref{eq:pois-2}), we get
the third compatibility condition or the NLSE,
\begin{equation}
i\frac{\partial\phi_{1,1}}{\partial\tau}+C_{1}\frac{\partial^{2}\phi_{1,1}}{\partial\xi^{2}}+C_{2}|\phi_{1,1}|^{2}\phi_{1,1}=C_{3}s,
\end{equation}
where
\begin{eqnarray}
C_{1} & = & -\frac{3}{2}kA\frac{\delta_{i}^{1/2}}{B^{5/2}},\\
C_{2} & = & -2\delta_{i}\frac{k^{4}}{\omega^{2}}+\frac{1}{2\omega^{2}}\left(T_{1}\delta_{i}^{1/2}+T_{2}+T_{3}\right),\\
C_{3} & = & -\frac{\omega^{3}}{2k^{2}\delta_{i}},
\end{eqnarray}
with
\begin{eqnarray}
T_{1} & = & 4\omega k\frac{A}{B^{3/2}}(Ak^{2}-D),\\
T_{2} & = & \omega^{2}(Ak^{2}-2D),\\
T_{3} & = & \frac{1}{B^{3}}\left(AD\frac{\omega}{k}\right)^{2}.
\end{eqnarray}

\bibliographystyle{unsrt}
\bibliography{cite}

\end{document}